\documentclass[11pt]{article}

\usepackage{times}
\usepackage{xspace}
\usepackage{graphicx}
\usepackage{amssymb}
\usepackage{amsthm}
\usepackage{fncylab}
\usepackage{fullpage}

 \newtheorem{theorem}{Theorem}
 \newtheorem{lemma}{Lemma}
 \newtheorem{corollary}{Corollary}
 \newtheorem{definition}{Definition}
\newtheorem{observation}{Observation}

\newcommand{\Alabel}[1]{\labelformat{myalgorithm}{Alg.~{\arabic{myalgorithm}}}\refstepcounter{myalgorithm}{\mbox{}\hfill{\fbox{\small\sc{}Alg.\,\arabic{myalgorithm}}}{\label{#1}}}}

\newcommand{\algfont}{\normalsize\rm}
\newcommand{\tab}{\hspace*{0.2in}}

\newcounter{myline}
\newcounter{myalgorithm}

\newenvironment{alg}{
  \par
  \algfont
    \noindent
    ~\begin{tabular}{|@{}|l|}\hline
      \begin{minipage}{0.97\textwidth}\raggedright
        \begin{list}{\arabic{myline}.}{
            \usecounter{myline}
            \setlength{\listparindent}{0in}
            \setlength{\topsep}{0in}
            \setlength{\itemsep}{0in}
            \setlength{\parsep}{0in}
            \setlength{\rightmargin}{0in}
            \setlength{\itemindent}{0in}
            \setlength{\labelsep}{0.035in}
            \setlength{\leftmargin}{0.18in}
          }
        \vspace*{0.06in}
          }{
        \vspace*{0.06in}
        \end{list}
      \end{minipage}\\\hline
    \end{tabular}
    \par
    \noindent
}
\newcommand{\A}{\item}

\newcommand{\Along}[1]{\item\parbox[t]{0.95\linewidth}{ #1}}
\newcommand{\Ahead}[1]{\item[]\hspace*{-\leftmargin}{#1}}
\newcommand{\Ain}[1]{\item[]\hspace*{-\leftmargin}{~\textrm{\small{\bf Input:} #1}}}
\newcommand{\Aout}[1]{\item[]\hspace*{-\leftmargin}{~\textrm{\small{\bf Output:} #1}}}

\newcommand{\algbeg}{%
  \addtolength{\labelsep}{0.15in}
  \addtolength{\itemindent}{0.15in}
  \addtolength{\listparindent}{0.15in}
  \addtolength{\linewidth}{-0.15in}
}
\newcommand{\algend}{%
  \addtolength{\labelsep}{-0.15in}
  \addtolength{\itemindent}{-0.15in}
  \addtolength{\linewidth}{0.15in}
}

\newcommand{\comment}[1]{\hfill \raisebox{0pt}{$\ldots$}\hspace{2pt}{\em  #1}}

\newcommand{\prob}[1]{{\sc\textscale{.955}{#1}}\xspace}
\makeatletter
\newcommand{\mymathfnnolimits}[1]{\ensuremath{{\mathop {\operator@font\sf #1}\nolimits }}}
\newcommand{\mymathfn}[1]{\ensuremath{\mathop {\operator@font\sf #1}}}
\makeatother
\newcommand{\eps}{\varepsilon}

\newcommand{\minimize}{\mymathfnnolimits{{minimize\,}}}
\newcommand{\maximize}{\mymathfnnolimits{{maximize\,}}}
\newcommand{\subjectto}{\mymathfnnolimits{{~subject~to~}}}

\newcommand{\giv}{\,|\,}
\newcommand{\opt}{\mbox{\sc opt}\xspace}

\newcommand{\deps}{\mymathfnnolimits{{vars}}}
\newcommand{\vars}{\deps}

\newcommand{\calA}{{\cal A}}

\newcommand{\calC}{{\cal C}}

\newcommand{\calF}{{\cal F}}

\newcommand{\calL}{{\cal L}}

\newcommand{\calR}{{\cal R}}

\newcommand{\R}{\mathbb{R}}
\newcommand{\Rp}{\R_{\ge 0}}
\newcommand{\Rpi}{\bar{\R}_{\ge 0}}

\newcommand{\Z}{\mathbb{Z}}
\newcommand{\Zp}{\Z_{\ge 0}}
\newcommand{\Zpi}{\bar{\Z}_{\ge 0}}

\newcommand{\mathspread}{}

\newcommand{\Spread}{\renewcommand{\mathspread}{\,}}
\newcommand{\SPREAD}{\renewcommand{\mathspread}{\,\,\,}}
\newcommand{\dospread}[1]{{\mathspread{#1}\mathspread}}
\Spread

\newcommand{\Eq}{\dospread{=}}

\newcommand{\Le}{\dospread{\le}}
\newcommand{\Ge}{\dospread{\ge}}

\newenvironment{Proof}{\begin{proof}}{\end{proof}}

\let\oldmarginpar\marginpar
\renewcommand\marginpar[1]{\-\oldmarginpar[\raggedleft\footnotesize #1]%
{\raggedright\footnotesize #1}}

\newcommand{\step}{\mymathfnnolimits{\sf step}\xspace}

\newcommand{\rstep}{\mymathfnnolimits{\sf random\_step}\xspace}

\newcommand{\degree}{\ensuremath{\Delta}\xspace}

\renewcommand{\prob}[1]{{\small\sc #1}}

\renewcommand{\opt}{\mymathfnnolimits{\sf opt}}

\newcommand{\cost}{\mymathfnnolimits{\sf cost}}
\newcommand{\size}{\mymathfnnolimits{\sf size}}
\newcommand{\conv}{\mymathfnnolimits{\sf conv}}

\newcommand{\cacheable}{\mymathfnnolimits{\sf cacheable}}

\newcommand{\rescp}[1]{{\tilde c'_{#1}}}
\newcommand{\resc}[1]{{\tilde c_{#1}}}
\newcommand{\rescost}[2]{\resc{#1}(#2)}
\newcommand{\rescostp}[2]{\rescp{#1}(#2)}

\newcommand{\xa}{x'}  
\newcommand{\xb}{x}  
\newcommand{\xhat}{\hat x}  
\newcommand{\xalg}{x^a}  

\begin{document}
\title{Greedy $\degree$-Approximation Algorithm for
Covering\\ with Arbitrary Constraints and Submodular Cost}
\author
{
	Christos Koufogiannakis \and Neal E.\ Young
\thanks{Department~of Computer Science and Engineering, University~of California, Riverside.}
}

\date{\today}

\maketitle

\begin{abstract}
This paper describes a simple greedy $\degree$-approximation algorithm
for any covering problem whose objective function is submodular and non-decreasing,
and whose feasible region can be expressed as the intersection of
arbitrary (closed upwards) covering constraints,
each of which constrains at most $\degree$ variables of the problem.
(A simple example is {\sc Vertex Cover}, with $\degree=2$.)
The algorithm generalizes
previous approximation algorithms for fundamental covering problems
and online paging and caching problems.
\end{abstract}

\setcounter{page}{1}
\newcommand{\Cite}[1]{{\,\scriptsize\cite{#1}}}
\newcommand{\Here}[1]{{\hfill\scriptsize[#1]}}

\newcommand{\TS}{\hspace{6pt}}

\section{Introduction and summary}
\label{sec:intro}\label{sec:introduction}

The classification of general techniques 
is an important research program
within the field of approximation algorithms.
What abstractions are useful for capturing a wide variety of problems and analyses?
What are the scopes of, and the relationships between,
the various algorithm-design techniques such as the primal-dual method, 
the local-ratio method \cite{Bar-Yehuda04Local}, and randomized rounding?
Which problems admit optimal and fast greedy approximation algorithms \cite{Khuller07Greedy,Borodin05How-well,Borodin11How-well}?
What general techniques are useful for designing online algorithms?
What is the role of locality among constraints and variables
\cite{papadimitriou1993linear,Kuhn06The-price,Bar-Yehuda04Local}?
We touch on these topics, exploring a simple greedy algorithm
for a general class of covering problems.
The algorithm has approximation ratio $\degree$
provided each covering constraint in the instance constrains only $\degree$ variables.

\smallskip
Throughout the paper, $\Rpi$ denotes $\Rp\cup\{\infty\}$
and $\Zpi$ denotes $\Zp\cup\{\infty\}$.

\smallskip
The conference version of this paper is \cite{koufogiannakis2009greedy}.
The journal version has been accepted to Algorithmica.

\begin{definition}[\bf Submodular-Cost Covering]
An instance is a triple $(c,\calC,U)$, where
\begin{itemize}
\item The cost function $c:\Rpi^n\rightarrow \Rpi$ is submodular,\footnote
{Formally, 
$c(x) + c(y) \ge c(x \wedge y) + c(x \vee y)$,
where $x\wedge y$ (and $x\vee y$)
are the component-wise minimum
(and maximum) of $x$ and $y$.
Intuitively, there is no positive synergy
between the variables:
let $\partial_j c(x)$ denote the rate at which increasing $x_j$ 
would increase $c(x)$;
then, increasing $x_i$ (for $i\neq j)$ does not increase $\partial_j c(x)$.
Any separable function $c(x) = \sum_j c_j(x_j)$ is submodular,
the product $c(x) = \prod_j x_j$ is not.
The maximum $c(x) = \max_j x_j$ is submodular,
the minimum $c(x) = \min_j x_j$ is not.}
continuous, and non-decreasing.

\smallskip

\item The constraint set $\calC\subseteq 2^{\Rpi^n}$ is a collection of covering constraints,
where each constraint $S\in\calC$ is a subset of $\Rpi^n$
that is closed upwards\footnote
{If $y\ge x$ and $x\in S$, then $y\in S$,
perhaps the minimal requirement for a constraint to be called a ``covering'' constraint.}
and under limit.
We stress that each $S$ may be non-convex.
\smallskip

\item
For each $j\in[n]$,
the domain $U_j$ (for variable $x_j$) is any subset of $\Rpi$ that is closed under limit.
\end{itemize}
The problem is to
find $x\in\Rpi^n$,
minimizing $c(x)$ subject to
$x_j \in U_j$\, ($\forall j\in[n]$)
and $x\in S$ ($\forall S\in\calC$).
\end{definition}

The definition assumes the objective function $c(x)$ 
is defined over all $x\in\Rpi^n$, 
even though the solution space is constrained to $x\in U$.
This is unnatural,
but any $c$ that is properly defined on $U$
extends appropriately\footnote
{One way to extend $c$ from $U$  to $\Rpi$: take the cost of $x\in\Rpi^n$
to be the expected cost of $\tilde x$,
where $\tilde x_j$ is rounded up or down 
to its nearest elements $a,b$ in $U_j$ such that $a\le x_j \le b$:
take $\tilde x_j = b$ with probability $\frac{b-x_j}{b-a}$,
otherwise take $\tilde x_j = a$.
If $a$ or $b$ doesn't exist, let $\tilde x_j$ be the one that does.
As long as $c$ is non-decreasing, sub-modular,
and (where appropriate) continuous over $U$, 
this extension will have these properties over $\Rpi^n$.
}
to $\Rpi^n$. 
In the cases discussed here
$c$ extends naturally to $\Rpi^n$
and this issue does not arise.

We call this problem \prob{Submodular-Cost Covering}.\footnote
{Changed from ``\prob{\footnotesize Monotone Covering}'' in the conference version 
 \cite{koufogiannakis2009greedy} due to name conflicts.}

\smallskip
For intuition, consider the well-known \prob{Facility Location} problem.
An instance is specified by a collection of customers, a collection of facilities, 
an opening cost $f_j\ge 0$ for each facility,
and an assignment cost $d_{ij}\ge 0$ for each customer $i$ and facility $j\in N(i)$.
The problem is to open a subset $\calF$ of the facilities so as to minimize
the cost to open facilities in $\calF$ 
(that is, $\sum_{j\in \calF} f_j$)
plus the cost for each customer to reach its nearest open, admissible facility
(that is, $\sum_{i} \min\{d_{ij} ~|~ j\in \calF\}$).
This is equivalent to \prob{Submodular-Cost Covering} instance $(c,\calC,U)$, with
\begin{itemize}
\item a variable $x_{ij}$ for each customer $i$ and facility $j$,
  with domain $U_{ij} = \{0,1\}$,
\item for each customer $i$, (non-convex) constraint $\max_{j\in N(i)} x_{ij}  \ge 1$
  (the customer is assigned a facility),
\item and (submodular) cost $c(x) = \sum_{j} f_j \max_{i} x_{ij} ~+~ \sum_{i,j} d_{ij} x_{ij}$
  (opening cost plus assignment cost).
\end{itemize}

\paragraph{A greedy algorithm for Submodular-Cost Covering (Section~\ref{sec:generic}).}
The core contribution of the paper is a greedy $\degree$-approximation algorithm for the problem, where $\degree$ is the maximum number of variables that any individual covering constraint $S$ in $\calC$ constrains.

For $S\in\calC$, let $\vars(S)$ contain the indices of variables that $S$ constrains
(i.e, $j\in\vars(S)$ if membership of $x$ in $S$ depends on $x_j$).
The algorithm is roughly as follows.  
\begin{quote}
{\em Start with an all-zero solution $x$,
then repeat the following step until all constraints are satisfied:
Choose any not-yet-satisfied constraint $S$.
To satisfy $S$,
raise each $x_j$ for $j\in\vars(S)$
(i.e., raise the variables that $S$ constrains),
so that each raised variable's increase contributes 
the same amount to the increase in the cost.}
\end{quote}
Section~\ref{sec:generic} gives the full algorithm and its analysis.

\paragraph{Fast implementations (Section~\ref{sec:implementation}).}
One important special case of \prob{Submodular-Cost Covering}
is \prob{Covering Integer Linear Programs} 
with upper bounds on the variables (\prob{CIP-UB}),
that is, problems of the form
\(\min\{c\cdot x ~|~ x\in \Zp^n;~ Ax \ge b; ~x\le u\}\)
where each $c_j$, $b_i$, and $A_{ij}$ is non-negative.
This is a \prob{Submodular-Cost Covering} instance  $(c,U,\calC)$
with variable domain $U_j = \{0,1,\ldots,u_j\}$ for each $j$
and a covering constraint $A_i x \ge b_i$ for each $i$,
and $\degree$ is the maximum number of non-zeros in any row of $A$.

Section~\ref{sec:implementation}
describes a nearly linear-time implementation for
a generalization of this problem:
\prob{Covering Mixed Integer Linear Programs}
with upper bounds on the variables
(\prob{CMIP-UB}).
As summarized in the bottom half of Fig.~\ref{fig:covering},
Section~\ref{sec:implementation}
also describes fast implementations
for other special cases:
\prob{Vertex Cover},
\prob{Set Cover},
\prob{Facility Location} (linear time);
and two-stage probabilistic \prob{CMIP-UB} (quadratic time).

\begin{figure}[t]
\noindent\centerline{\small
\begin{tabular}[t]{|@{\TS}l@{\TS\TS}l@{\TS}l@{\TS\TS}l@{\TS}|} \hline
problem & approximation ratio & method \hfill where & comment \\ \hline
\prob{Vertex Cover} & \rule{0em}{1.1em}$2- \ln \ln\widehat\degree/\ln\widehat\degree$ & 
local ratio
\hfill\Cite{Halperin2002Improved}&
see also
\Cite{Hochbaum83BoundsSetCover,Bar-Yehuda85A-local-ratio,Monien1985RamseyNumbers,Halldorsson1994GreedIsGood,Halldorsson1997GreedIsGood,Hastad2001SomeOptional,Dinur2005OnTheHardness,Khot2008VertexCover}
\\
\prob{Set Cover} \hfill  & $\degree$ & LP; greedy \hfill
\Cite{Hochbaum82Approximation,Hochbaum83BoundsSetCover};
\Cite{Bar-Yehuda81A-Linear-Time}
& $\degree=\max_i|\{j ~|~ A_{ij}>0\}|$ 
\hfill $\star$\\
CIP-01 w/$A_{ij}\in\Zp$ \hfill   &  $\max_i \sum_j A_{ij}$ & primal-dual \hfill \Cite{Bertsimas98Rounding,Hall86A-fast}
& quadratic time
\hfill $\star$\\
CIP-UB \hfill  & $\degree$ & ellipsoid  \hfill\Cite{Carr00Strengthening,Pritchard09Approximability,Pritchard11Approximability}
& KC-ineq., high-degree-poly time
\hfill $\star$
\\ \hline \hline 
& & & \\[-8pt]
\prob{Submod-Cost Cover} & $\degree$ & greedy \Here{our \bf \S\ref{sec:generic}} & $\min\{ c(x) ~|~ x\in S~(\forall S\in\calC)\}$ \hfill \small new
\\
\prob{Set/Vertex Cover} & $\degree$  & greedy \Here{our \bf \S\ref{sec:facility_location}} & linear time 
\\
\prob{Facility Location} & $\degree$  & greedy \Here{our \bf \S\ref{sec:facility_location}} & linear time 
\hfill \small new\\
\prob{CMIP-UB} & $\degree$  & greedy \Here{our \bf \S\ref{sec:restriction}} & near-linear time 
\hfill \small new\\
\prob{2-stage CMIP-UB} & $\degree$  & greedy \Here{our \bf \S{\ref{sec:submodular}}} & quadratic time 
\hfill \small new \\ \hline
\end{tabular}
}
\caption{Some $\degree$-approximation algorithms for covering problems.
``$\star$'' = generalized or strengthened here.
}\label{fig:covering}
\end{figure}

\paragraph{Related work: $\degree$-approximation algorithms for classical covering problems  (top half of Fig.~\ref{fig:covering}).}
See e.g. \cite{hochbaum1996aan,vazirani2001aa} for an introduction to classical covering problems.
For \prob{Vertex Cover}\footnote
{
\prob{\footnotesize Set Multicover} 
is \prob{\footnotesize CIP-UB} restricted to $A_{ij}\in\{0,1\}$;
\prob{\footnotesize Set Cover} 
is \prob{\footnotesize Set Multicover} restricted to $b_i=1$;
\prob{\footnotesize Vertex Cover} 
is \prob{\footnotesize Set Cover} restricted to $\degree=2$.
}
and \prob{Set Cover}
in the early 1980's, Hochbaum gave a $\degree$-approximation algorithm
based on rounding an LP relaxation
\cite{Hochbaum82Approximation};
Bar-Yehuda and Even gave a linear-time greedy algorithm
(a special case of the algorithms here)
\cite{Bar-Yehuda81A-Linear-Time}.
A few years later
Hall and Hochbaum gave a quadratic-time primal-dual algorithm
for \prob{Set Multicover} \cite{Hall86A-fast}.
In the late 1990's, Bertsimas and Vohra further generalized that result
with a quadratic-time primal-dual algorithm
for \prob{Covering Integer Programs}
with $\{0,1\}$-variables 
 (\prob{CIP-01}),
but restricted to integer constraint matrix $A$
and with approximation ratio $\max_i\sum_j A_{ij} \ge \degree$
\cite{Bertsimas98Rounding}.
In 2000,  Carr et al.\ 
gave the first $\degree$-approximation for \prob{CIP-01} \cite{Carr00Strengthening}.
In 2009 (independently of our work), Pritchard
extended that result to \prob{CIP-UB}
\cite{Pritchard09Approximability,Pritchard11Approximability}.
Both \cite{Carr00Strengthening} and \cite{Pritchard09Approximability,Pritchard11Approximability}
use the (exponentially many) Knapsack-Cover (KC) inequalities
to obtain integrality gap\footnote
{The standard LP relaxation has arbitrarily large gap
(e.g.~$\min \{x_1 ~|~ 10 x_1 + 10 x_2 \ge 11; x_2 \le 1\}$ has gap 10).

\cite{Carr00Strengthening} states (without details)
that their CIP-01 result extends CIP-UP,
but it is not clear how (see \cite{Pritchard09Approximability,Pritchard11Approximability}).}
$\degree$,
and their algorithms use the ellipsoid method,
so have high-degree-polynomial running time.

As far as we know, 
\prob{Set Cover} is the most general special case of \prob{Submodular-Cost Covering}
for which any nearly linear time $\degree$-approximation algorithm was previously known,
while \prob{CIP-UB} is the most general special case
for which any polynomial-time $\degree$-approximation algorithm was previously known.

Independently of this paper, Iwata and Nagano give $\degree$-approximation 
algorithms for variants of \prob{Vertex Cover}, \prob{Set Cover}, and \prob{Edge Cover} with
submodular (and possibly decreasing!) cost \cite{iwata2009submodular}.

\paragraph{Online covering, paging, and caching (Section~\ref{sec:online}).}
In {\em online} covering
(following, e.g.~\cite{Buchbinder05OnLine,Buchbinder09Online,Bansal07A-Primal-Dual}),
the covering constraints are revealed one at a time in any order.
An online algorithm must choose an initial solution $x$,
then, as each constraint ``$x\in S$'' is revealed, must increase variables in $x$
to satisfy the constraint, without knowing future constraints and without decreasing any variable.
The algorithm has {\em competitive ratio} $\degree$ if the cost of its final solution
is at most $\degree$ times the optimal (offline) cost (plus a constant
that is independent of the input sequence).
The algorithm is said to be {\em $\degree$-competitive}. 

The greedy algorithm here
is a $\degree$-competitive online algorithm
for \prob{Submodular-Cost Covering}.

As recently observed in~\cite{Buchbinder05OnLine,Buchbinder09Online,Bansal07A-Primal-Dual},
many classical online paging and caching problems reduce to online covering
(usually online \prob{Set Cover}).
Via this reduction, the algorithm here generalizes 
many classical deterministic online paging and caching algorithms.
\begin{figure}[t]
\noindent\centerline{\small
\begin{tabular}[t]{|@{\TS}l@{\TS\TS}l@{\TS}l@{\TS\TS}l@{\TS}|} \hline
online problem & competitive ratio & deterministic online  &  comment \\ \hline
\prob{ski rental}\hfill & 2; $\frac{e}{e-1}$ & det.; random \hfill\Cite{Karlin1988CompetitiveSnoopy,Lotker08RentLeaseBuy} & \hfill $\star$
\\
\prob{paging} \hfill & $k=\degree$ & 
potential function
\hfill\Cite{Sleator85Amortized,Raghavan94Memory} 
&  e.g.  LRU, FIFO, FWF, Harmonic\hfill $\star$
\\
\prob{connection caching}  & $O(k)$ &
reduction to paging \hfill\Cite{Cohen1999ConnectionCaching,Albers02On-Generalized} &  
 \hfill $\star$
\\
\prob{weighted caching} \hfill & $k$ & primal-dual\hfill\Cite{young1991online,Young94The-k-Server,Raghavan94Memory} 
&  e.g.  Harmonic, Greedy-Dual \hfill $\star$
\\
\prob{file caching} \hfill & $k$ & primal-dual \hfill\Cite{Young98Online,Young02On-Line,Cao97Cost-aware} 
&  e.g.  Greedy-Dual-Size, Landlord\hfill $\star$
\\
\prob{unw.~set cover} & $O(\log\degree \log \frac{n}{\opt})$ & primal-dual \hfill\Cite{Buchbinder05OnLine,Buchbinder09Online}
& unweighted
\\
\prob{CLP} & $O(\log n)$ & fractional \hfill\Cite{Buchbinder05OnLine,Buchbinder09Online}
&$\min\{ c\cdot x ~|~ Ax \ge b; x \le u\}$,
\\ \hline \hline
& & & \\[-7pt]
\prob{Submod-Cost Cover} & $\degree$ & 
potential function
\Here{our \bf \S\ref{sec:generic}} & 
includes the above, CMIP-UB~
\hfill \small new
\\\prob{upgradable caching} & $d+k$ & 
by reduction 
\Here{our \bf \S\ref{sec:online}} & 
$d$ components, $k$ files in cache
\hfill \small new
\\ \hline
\end{tabular}
}
\caption{$\degree$-competitive online paging and caching.
``$\star$'' = generalized or strengthened here.
}\label{fig:online}
\end{figure}
These include \prob{LRU} and \prob{FWF} for \prob{Paging} \cite{Sleator85Amortized},
\prob{Balance} and \prob{Greedy Dual} for \prob{Weighted Caching}
\cite{Chrobak91New-results,young1991online,Young94The-k-Server},
\prob{Landlord}
\cite{Young98Online,Young02On-Line} (a.k.a.~\prob{Greedy Dual Size}) \cite{Cao97Cost-aware},
for \prob{File Caching},
and algorithms for \prob{Connection Caching}
\cite{Cohen1999ConnectionCaching,Cohen2000ConnectionCaching,Cohen2003ConnectionCaching,Albers02On-Generalized}
(all results marked with ``$\star$'' in Fig.~\ref{fig:online}).

As usual, the competitive ratio $\degree$ is the cache size, commonly denoted $k$,
or, in the case of \prob{File Caching}, the maximum number of files
ever held in cache (which is at most the cache size).

Section~\ref{sec:online} illustrates this connection
using \prob{Connection Caching} as an example.

Section~\ref{sec:online} also illustrates the generality
of online \prob{Submodular-Cost Covering} by describing a $(d+k)$-competitive algorithm
for a new class of {\em upgradable} caching problems,
in which the online algorithm chooses not only which
pages to evict, but also how to pay to upgrade $d$ 
hardware parameters (e.g.~cache size, CPU, bus, network speeds, etc.)
to reduce later costs and constraints 
(somewhat like \prob{Ski Rental} 
\cite{Karlin1988CompetitiveSnoopy} 
and multi-slope \prob{Ski Rental} \cite{Lotker08RentLeaseBuy} ---
special cases of online \prob{Submodular-Cost Covering} with $\degree=2$).

Section~\ref{sec:random} describes 
a natural randomized generalization of the greedy algorithm
(\ref{alg:submodular}),
with even more flexibility in incrementing the variables.
This yields a {\em stateless} $\degree$-competitive 
online algorithm for \prob{Submodular-Cost Covering},
generalizing 
Pitt's \prob{Vertex Cover} algorithm \cite{Bar-Yehuda00One-for-the-price}
and the \prob{Harmonic} $k$-server algorithm 
as it specializes for \prob{Paging} and \prob{Weighted Caching}
\cite{Raghavan94Memory}.

\paragraph{Related work: randomized online algorithms.}
For most online problems here, no deterministic online algorithm
can be better than $\degree$-competitive (where $\degree=k$),
but better-than-$\degree$-competitive {\em randomized} online algorithms
are known.
Examples include \prob{Ski Rental}
\cite{Karlin1988CompetitiveSnoopy,Lotker08RentLeaseBuy},
\prob{Paging}  \cite{Fiat91Competitive,Mcgeoch91AStrongly},
\prob{Weighted Caching} \cite{Bansal07A-Primal-Dual,Cao97Cost-aware},
\prob{Connection Caching} \cite{Cohen1999ConnectionCaching},
and \prob{File Caching} \cite{Bansal08RandomizedCompetitive}.
Some cases of online \prob{Submodular-Cost Covering}
(e.g.~\prob{Vertex Cover})
are unlikely to have better-than-$\degree$-competitive randomized algorithms.
It would be interesting to classify which cases
admit better-than-$\degree$-competitive randomized online algorithms.

\paragraph{Relation to local-ratio and primal-dual methods (Section~\ref{sec:relation}).}

Section~\ref{sec:relation} describes how
the analyses here can be recast (perhaps at some expense in intuition)
in either the local-ratio framework or (at least for linear costs)
the primal-dual framework.
Local ratio is usually applied to problems with variables in $\{0,1\}$;
the section introduces an interpretation of local ratio for more general domains,
based on residual costs.

Similarly, the Knapsack Cover (KC) inequalities are most commonly used
for problems with variables in $\{0,1\}$,
and it is not clear how to extend the KC inequalities to more general domains
(e.g.~from \prob{CMIP-01} to \prob{CMIP-UB}).
(The standard KC inequalities suffice for $O(\log(\widehat\degree))$-approximation 
of \prob{CMIP-UB} \cite{Kolliopoulos05Approximation},
but may require some modification to give $\degree$-approximation
of \prob{CMIP-UB} \cite{Pritchard09Approximability,Pritchard11Approximability}.)
The primal-dual analysis in Section~\ref{sec:relation} 
uses a new linear program (LP) relaxation for \prob{Linear-Cost Covering}
that may help better understand how to extend the KC inequalities.

Section~\ref{sec:relation} also discusses how
the analyses here can be interpreted 
via a certain class of valid linear inequalities,
namely inequalities that are ``local'' in that
they can be proven valid by looking only at each single
constraint $S\in\calC$ in isolation.

\paragraph{Related work: hardness results, log-approximation algorithms.}
Even for simple covering problems such as \prob{Set Cover},
no polynomial-time $(\degree-\eps)$-approximation algorithms (for any constant $\eps>0$) 
are currently known
for small (e.g.~constant) $\degree$.
A particularly well studied special case, with $\degree=2$,
is \prob{Vertex Cover},  for which some complexity-theoretic evidence suggests
that such an algorithm may not exist
\cite{Halperin2002Improved,Hochbaum83BoundsSetCover,Bar-Yehuda85A-local-ratio,Monien1985RamseyNumbers,Halldorsson1994GreedIsGood,Halldorsson1997GreedIsGood,Hastad2001SomeOptional,Dinur2005OnTheHardness,Khot2008VertexCover}.

For instances where $\degree$ is large, $O(\log \widehat \degree)$-approximation algorithms
may be more appropriate, where $\widehat\degree$ is the maximum number of constraints
in which any variable occurs.
Such algorithms exist for \prob{set cover}
\cite[(greedy, 1975)]{Johnson73SetCover,Johnson74SetCover,Lovasz75SetCover,Chvatal79GreedySetCover}
for \prob{CIP}
\cite[(ellipsoid, 2000)]{Srinivasan99Improved,Srinivasan01NewApproaches}
and \prob{CIP-UB}
\cite[(ellipsoid/KC inequalities, 2005)]{Kolliopoulos05Approximation}.
It is an open question 
whether there is a fast {\em greedy} $O(\log\widehat\degree)$-approximation algorithm 
handling all of these problems (via, say, \prob{CIP-UB}).

Recent works with log-approximations for submodular-cost covering problems include
\cite{Hayrapetyan2005NetworkDesign,ravi2006hua,shmoys2004soa,Chudak07Efficient}.
Most of these have high-degree-polynomial run time.
For example, the $(\ln n)$-approximation algorithm
for two-stage probabilistic \prob{Set-Cover}
\cite{Hayrapetyan2005NetworkDesign}
requires solving instances of 
\prob{Submodular Function Minimization}
\cite{Orlin2007AFasterStrongly,Orlin2009AFasterStrongly},
which requires high-degree-polynomial run time.
(\cite{Hayrapetyan2005NetworkDesign}
also claims a related 2-approximation for two-stage probabilistic \prob{Vertex Cover} without details.)

\paragraph{Related work: distributed and parallel algorithms.}
Distributed and parallel approximation algorithms for covering problems 
are an active area of study.
The simple form of the greedy algorithm here makes it particularly amenable 
for distributed and/or parallel implementation.
In fact, it admits poly-log-time distributed and parallel implementations,
giving (for example) the first poly-log-time 2-approximation algorithms
for the well-studied (weighted) \prob{Vertex Cover} and \prob{Maximum Weight Matching} problems.
See \cite{Koufogiannakis09Distributed,Koufogiannakis09DistributedPacking,Koufogiannakis11Distributed}
for details and related results.


\paragraph{Organization.}
Section~\ref{sec:generic} gives the 
greedy algorithm for \prob{Submodular-Cost Covering} (\ref{alg:submodular})
and proves that is has approximation ratio $\degree$.
Section \ref{sec:online} describes applications to online problems. 
Section \ref{sec:random} describes randomized generalizations of the greedy algorithm,
including a stateless online algorithm.
Sections \ref{sec:local} and \ref{sec:relation}
explain how to view the analysis via the local-ratio and primal-dual methods.
Section \ref{sec:implementation} details fast implementations for some special cases.
After Section~\ref{sec:generic}, each section may be read independently of the others.

\section{Greedy Algorithm for Submodular-Cost Covering  (\ref{alg:submodular})}
\label{sec:generic}

This section gives the full greedy algorithm for \prob{Submodular-Cost Covering}
(\ref{alg:submodular}) and the analysis of its approximation ratio.
We assume
\prob{Submodular-Cost Covering} instances are given in {\em canonical form}:
\begin{definition}[\bf canonical form]\label{def:canonical}
An instance $(c,U,\calC)$ is in {\em canonical form}
if each variable domain is unrestricted (each $U_j=\Rpi$).
Such an instance is specified by just the pair $(c,\calC)$.
\end{definition}

This assumption is without loss of generality by the following reduction:
\begin{observation}\label{obs:canonical}
For any \prob{Submodular-Cost Covering} instance $(c,U,\calC)$,
there is an equivalent canonical form instance $(c,\calC')$.
By ``equivalent'', we mean that 
any $x$ that is feasible in $(c,U,\calC)$ is also feasible in $(c,\calC')$,
and that any $x'$ that is minimally feasible in $(c,\calC')$ 
is also feasible in $(c,U,\calC)$.
\end{observation}
\noindent
Given any feasible solution $x'$ to $(c,\calC')$,
one can compute a feasible solution $x$ to $(c,U,\calC)$ with $c(x)\le c(x')$
by taking each $x_j = \max\{\alpha\in U_j ~|~ \alpha \le x_j\}$.

The reduction is straightforward and is given in the Appendix.
The idea is to incorporate the variable-domain restrictions ``$x_j\in U_j$''
directly into each covering constraint $S\in \calC$,
replacing each occurrence of $x_j$ in each $S$ by $\max\{\alpha\in U_j ~|~ \alpha\le x_j\}$.
For example, applied to a \prob{CIP-UB} instance $(c,U,\calC)$
as described in the introduction,
the reduction produces the canonical instance $(c,\calC')$
in which each covering constraint $A_i x\ge b_i$ in $\calC$
is replaced in $\calC'$ by the stronger non-convex covering constraint
\[\textstyle
\sum_j A_{ij} \lfloor \min(x_j,u_j) \rfloor ~\ge~ b_i.
\]
To satisfy these constraints,
it doesn't help to assign any $x_j$ a value outside of $U_j=\{0,1,\ldots,u_j\}$:
any minimal $x$ satisfying the constraints in $\calC'$
will also satisfy $x_j\in\{0,1,\ldots,u_j\}$ for each $j$.

In the rest of the paper, we assume all instances are given in canonical form.
To handle an instance $(c,U,\calC)$ that is not in canonical form,
apply the above reduction to obtain canonical instance $(c,\calC')$,
use one of the algorithms here to compute a $\degree$-approximate
solution $x$ for $(c,\calC')$, then compute vector $x'$ 
as described after Observation~\ref{obs:canonical}.

\newcommand{\inX}[1]{_{\!\!{}_{\!{#1}}\,\,}}
\newcommand{\inS}{{\inX S}}

\begin{definition}
For any covering constraint $S$ and $x\in\Rp^n$,
let ``$x\le\inS y$'', ``$x >\inS y$'', etc., 
mean that the operator holds coordinate-wise for coordinates in $\vars(S)$.
E.g.~$x \le\inS y$ if $x_j\le y_j$ for all $j\in\vars(S)$.
\end{definition}
\begin{observation}\label{obs:up}
If $x \in S$ and $y\ge\inS x$, then $y\in S$.
\end{observation}
The observation is true simply because $S$ is closed upwards,
and membership of $y$ in $S$ depends only on $y_j$ for $j\in\vars(S)$.
We use this observation throughout the paper.

To warm up the intuition for \ref{alg:submodular},
we first introduce and analyze a simpler version, \ref{alg:simple},
that works only for linear costs.
The algorithm starts with $x\leftarrow \mathbf 0$,
then repeatedly chooses any unmet constraint $S$,
and, to satisfy $S$,
raises all variables $x_j$ with $j\in\vars(S)$ at rate $1/c_j$, until $x$ satisfies $S$:

\medskip

\begin{alg}
\Ahead{\bf\small Greedy algorithm for Linear-Cost Covering} 
\Alabel{alg:simple}
\Ain{(linear) cost vector $c\in\Rp^n$, canonical constraints $\calC$}
\Aout{$\degree$-approximate solution $x$}
\A Recall that $\vars(S)$ contains the indices of variables that $S$ constrains.
\A Start with $x\leftarrow \mathbf 0$, then, for each of the constraints $S\in\calC$, in any order:
\algbeg
\A Just until $x\in S$, do:
\A \tab for all $j\in\vars(S)$ simultaneously, raise $x_j$  continuously at rate $1/c_j$. \label{line:key}
\algend
\A Return $x$.  
\end{alg}

\noindent
As the variables increase in Line~\ref{line:key},
the cost of $x$ increases at rate $|\vars(S)|\le\degree$
(each variable contributes to the cost increase at unit rate).\footnote
{If some $c_j=0$, then $x_j$ is raised instantaneously to $\infty$ at cost 0,
after which the cost of $x$ increases at rate less than $|\vars(S)|$.}
 The proof of the approximation ratio relies on the following observation:
\begin{observation}\label{obs:under}
Let $y$ be any feasible solution.   
Consider an iteration for a constraint $S$.
Unless the current solution $x$ already satisfies $S$ at the start of the iteration,
at the end of the iteration,
$x$ has some variable $x_k$ with $k\in\vars(S)$
such that $x_k\le y_k$.  (That is, $x\not>\inS y$.)
\end{observation}
\begin{Proof} 
At the start of the iteration, since $y$ but not $x$ satisfies $S$,
Observation~\ref{obs:up} implies that $x\not\ge\inS y$. 
During the iteration, while Line~\ref{line:key} is raising the $x_j$ for $j\in\vars(S)$, 
if at some moment $x\ge\inS y$,  then, since $y\in S$,
it must be (by Observation~\ref{obs:up}) that $x\in S$ also,
so at that moment Line~\ref{line:key} stops raising the variables,
before $x >\inS y$.
\end{Proof}

As the variables increase in Line~\ref{line:key},
Observation~\ref{obs:under} implies that, for some $x_k$,
the growing interval $[0,x_k]$ covers (at rate $1/c_k$)
a larger and larger fraction of the corresponding interval $[0,x^*_k]$ 
in the optimal solution $x^*$.
This allows the $\degree$-rate increase in the cost of $x$ 
to be charged at unit rate to the cost of $x^*$,
proving that the final cost of $x$ 
is at most $\degree$ times the cost of $x^*$.

\begin{figure}[ht]
  \centering
  \includegraphics[width=0.8\textwidth]{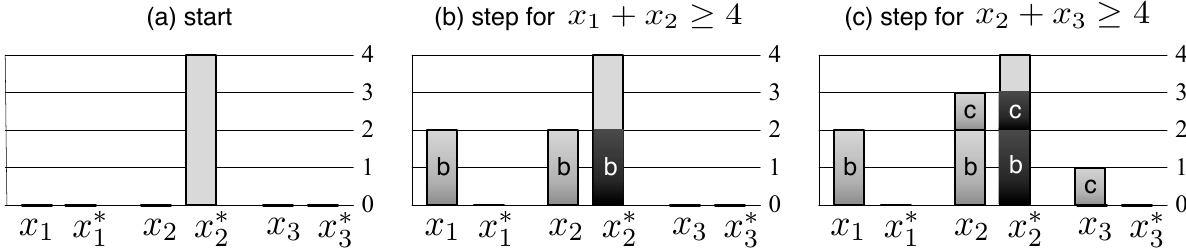}
 \caption{Two steps of \ref{alg:simple}, where $x^*=(0,2,0)$.
    Dark portions of $x^*_2$ have been charged.}
\label{fig:charging}
\end{figure}

For example, consider two iterations of \ref{alg:simple}
on input $\min\{x_1+x_2+x_3~|~ x_1+x_2\ge 4;~ x_2+x_3\ge 4\}$
with optimal solution $x^*=(0,4,0)$,
as shown in Fig.~\ref{fig:charging}.
The first iteration does a step for the first constraint,
raising $x_1$ and $x_2$ by 2,
and charging the cost increase of 4 to the $[0,2]$ portion of $x_2^*$.
The second iteration does a step for the second constraint,
raising $x_2$ and $x_3$ both by 1,
and charging the cost increase of 2 to the $[2,3]$ portion of $x_2^*$.

We generalize this charging argument
by defining the {\em residual problem for the current $x$},
which is the problem of finding a minimum-cost
augmentation of the current $x$ to make it feasible.
For example, after the first iteration of \ref{alg:simple} in the example above,
the residual problem for $x=(2,2,0)$ is equivalent to
$\min\{y_1+y_2+y_3~|~ y_1+y_2\ge 0;~ y_2+y_3\ge 2\}$.
For notational simplicity, in the definition of the residual problem,
instead of shifting each constraint, 
we (equivalently, but perhaps less intuitively) leave the constraints alone
but modify the cost function
(we charge $y$ only for the part of $y$ that exceeds $x$):~\footnote
{Readers may recognize a similarity to the local-ratio method.
 This is explored in Section~\ref{sec:local}.}
\begin{definition}[\bf residual problem]\label{def:residual}
Given any \prob{Submodular-Cost Covering} instance $(c,\calC)$,
and any $x\in\Rpi^n$, define the {\em residual problem for $x$} to be
the instance $(\resc x, \calC)$
with cost function $\rescost x y = c(x\vee y)-c(x)$.

For $Q\subseteq \Rp^n$, define the {\em cost of $Q$ in the residual problem for $x$} to be 
\(\rescost x Q = \min_{y\in Q} \rescost x y\).

If $Q$ is closed upwards, then \(\rescost x Q\) equals $\min \{c(y) - c(x) ~|~ y\ge x,~ y\in Q\}$.
\end{definition}
\noindent
In all cases here $Q$ is closed upwards,
and we interpret $\rescost x Q$ as
{\em the minimum increase in $c(x)$ necessary to raise coordinates of $x$ to bring $x$ into $Q$}.
The residual problem
$(\resc x,\calC)$ has optimal cost $\rescost x {\bigcap_{S\in\calC} S}$.

Here is the formal proof of the approximation ratio,
as it specializes for \ref{alg:simple}.

\begin{lemma}[\bf correctness of \ref{alg:simple}]\label{lemma:simple}
\ref{alg:simple} is a $\degree$-approximation algorithm for \prob{Linear-Cost Covering}.
\end{lemma}
\begin{Proof}
First consider the case when every cost $c_j$ is non-zero.
Consider an iteration for a constraint $S\in\calC$.

Fix any feasible $y$.
The cost $\rescost x y$ of $y$ in the residual problem for $x$ is
the sum $\sum_{j} c_j \max(y_j - x_j, 0)$.
As Line~\ref{line:key} raises each variable $x_j$ for $j\in\vars(S)$ at rate $1/c_j$,
by Observation~\ref{obs:under}, one of the variables being raised 
is an $x_k$ such that $x_{k} < y_k$.
For this $k$, the term $c_k \max(y_k-x_k, 0)$ in the sum
is decreasing at rate 1.
No terms in the sum increase.
Thus, $\rescost x y$ decreases at rate at least 1.

Meanwhile, the cost $c(x)$ of $x$ increases at rate $|\vars(S)|\le \degree$.
Thus, the algorithm maintains the invariant $c(x)/\degree + \rescost x y \le c(y)$
(true initially because $c(\mathbf 0) = 0$ and $\rescost {\mathbf 0} y = c(y)$).
Since $\rescost x y \ge 0$, this implies that $c(x) \le \degree c(y)$ at all times.

In the case that some $c_j=0$ during an iteration, 
the corresponding $x_j$'s are set instantaneously to $\infty$.
This increases neither $c(x)$ nor $\rescost x y$, so the above invariant
is still maintained and the conclusion still holds.
\end{Proof}

The main algorithm (\ref{alg:submodular}, next)
generalizes \ref{alg:simple} in two ways:
First, the algorithm works with any submodular (not just linear) cost function.
(This generalization is more complicated but technically straightforward.)
Second, in each iteration, instead of increasing variables just until the constraint is satisfied,
it chooses a {\em step size} $\beta\ge 0$ explicitly 
(we will see that this will allow a {\em larger} step than in \ref{alg:simple}). 
Then, for each $j\in\vars(S)$,
it increases $x_j$ maximally so that the cost $c(x)$ of $x$ increases by (at most) $\beta$.

\noindent
\begin{alg}
\Ahead{\bf Greedy algorithm for Submodular-Cost Covering} 
\Alabel{alg:submodular}
 \Ain{objective $c$, canonical constraints $\calC$}
\Aout{$\degree$-approximate solution $x$
(provided conditions of Thm.~\ref{thm:generic} are met).}
\A Let $x\leftarrow \mathbf 0$.  
\A While $\exists$ $S\in \calC$ such that $x\not\in S$ do:
\algbeg
\A Choose any $S$ such that $x\not\in S$ and do {\bf step}$_c(x,S)$ (defined below).
\algend
\A Return $x$.  
\Ahead{\bf {\small Subroutine}  step$_c$}
\comment{makes progress towards satisfying $x\in S$.}
\Ain{current solution $x$, unsatisfied $S\in \calC$}
\setcounter{myline}{0}
\A Choose any {\em step size} $\beta\in [0,\rescost x S]$.
\comment{discussed before Thm.~\ref{thm:generic}.}
\label{step:beta}
\Along{For each $j\in\vars(S)$,
let $\xa_j\in\Rpi$ be the maximum such~that raising $x_j$ to $\xa_j$ would increase $c(x)$ by at most $\beta$.
\comment{recall $c(x)$ is continuous}}
\label{step:choose}
\A For $j\in\vars(S)$, let $x_j\leftarrow  \xa_j$.

\label{step:raise}
\end{alg}

\paragraph{Choosing the step size $\mathbf{\beta}$.}
In an iteration for a constraint $S$, the algorithm can choose any step size $\beta\ge 0$
subject to the restriction
$\beta \le \rescost x S = \min\{ c(y) - c(x) ~|~ y\in S, ~y\ge x\}$.
That is, $\beta$ is at most the minimum cost that would be necessary
to increase variables in $x$ to bring $x$ into $S$.
To understand ways in which \ref{alg:submodular} can choose $\beta$, consider the following.
\begin{itemize}
\item In all cases, \ref{alg:submodular} can take $\beta$ as \ref{alg:simple} does:
just large enough to ensure $x\in S$ after the iteration.
By an argument similar to the proof of Lemma~\ref{lemma:simple}, this particular $\beta$ is guaranteed to satisfy the restriction $\beta\le \rescost x S$.
(Of course another option is to take any $\beta$ smaller than this $\beta$.)
\smallskip

\item In some cases, \ref{alg:submodular} can take $\beta$ larger than \ref{alg:simple} does.
For example, 
consider a linear constraint $x_u + x_w \ge 1$ with linear cost $c(x) = x_u + x_w$.
Consider an iteration for this constraint, starting with $x_u = x_w = 0$.
\ref{alg:simple} would take $\beta=1/2$ and $x_u=x_w=1/2$,
satisfying the constraint.
But $\rescost x S = 1$ (to bring $x$ into $S$ would require raising $x_u+x_w$ to 1),
so \ref{alg:submodular} can take $\beta=1$ and $x_u=x_w=1$,
``over-satisfying'' the constraint.
\smallskip

\item It would be natural to set $\beta$ 
to its maximum allowed value $\rescost x S$,
but this value can be hard to compute.
Consider a single constraint $S$:
$\sum_{j} c_j \min(1,\lfloor x_j\rfloor) \ge 1$,
with cost function $c(x) = \sum_j c_j x_j$.
Then $\rescost {\mathbf 0} S=1$
if and only if there is a subset $Q$
such that $\sum_{j\in Q} c_j = 1$.
Determining this for arbitrary $c$ is \prob{Subset Sum}, which is NP-hard.
Still, determining a ``good enough'' $\beta$ is not hard:
take, e.g.~$\beta = \min\{c_j(1-x_j)~|~x_j < 1\}$.
If $x\not\in S$, then this is at most $\rescost x S$
because to bring $x$ into $S$ would require raising at least one $x_j<1$ to 1.
This choice of $\beta$ is easy to compute,
and with it \ref{alg:submodular} will satisfy $S$ within $\degree$ iterations.
\end{itemize}
In short, computing $\rescost x S$ can be hard,
but finding a ``good'' $\beta \le \rescost x S$ is not hard.
A generic choice is to take $\beta$ just large enough 
to bring $x$ into $S$ after the iteration, as \ref{alg:simple} does,
but in some cases (especially in online,
distributed, and parallel settings where the algorithm is restricted)
other choices may be easier to implement or lead to fewer iterations.
For a few examples, see the specializations of \ref{alg:submodular}
in Section~\ref{sec:implementation}.

\smallskip 
The proof of the approximation ratio for \ref{alg:submodular}
generalizes the proof of Lemma~\ref{lemma:simple} in two ways:
the proof has a second case to handle step sizes $\beta$ larger
than \ref{alg:simple} would take,
and the proof handles the more general (submodular) cost function
(the generality of which makes this proof unfortunately more abstract).

\begin{theorem}[\bf correctness of \ref{alg:submodular}]\label{thm:generic}
For \prob{Submodular-Cost Covering},
if \ref{alg:submodular}
terminates, it returns a $\degree$-approximate solution.
\end{theorem}
\begin{Proof}
Consider an iteration for a constraint $S\in \calC$.
By the submodularity of $c$,
the iteration increases the cost $c(x)$ of $x$ by at most $\beta|\vars(S)|$. \footnote
{To see this, consider each variables $x_j$ for $j\in\vars(S)$ one at a time,
in at most $\degree$ steps;
by submodularity of $c$, 
in a step that increases a given $x_j$, the increase in $c(x)$ 
is at most what it would have been if $x_j$ had been increased first,
i.e., at most $\beta$.}
We show that, for any feasible $y$,
the cost $\rescost x y$ of $y$ in the residual problem for $x$
decreases by at least $\beta$.
Thus, the invariant $c(x)/\degree + \rescost x y \le c(y)$,
and the theorem, hold.

Recall that $x\wedge y$ (resp.~$x\vee y$)
denotes the coordinate-wise minimum (resp.~maximum) of $x$ and $y$.

Let $\xb$ and $\xa$ denote the vector $x$ before and after the iteration, respectively.
Fix any feasible $y$.  

First consider the case when $y\ge \xb$ (the general case will follow from this one).
The submodularity of $c$ implies
\(
c(\xa) + c(y) \ge c(\xa \vee y) + c(\xa\wedge y).
\)
Subtracting $c(\xb)$ from both sides and rearranging terms gives
(with equality if $c$ is separable, e.g.~linear)
\[
[c(y) - c(\xb)] \,-\, [c(y \vee \xa) - c(\xa)]  ~\ge~  c(\xa \wedge y) - c(\xb).
\]
The first bracketed term is $c(y\vee \xb) - c(\xb) = \rescost {\xb} y$
(using here that $y\ge \xb)$ so $y\vee \xb = y$).
The second bracketed term is $\rescost {\xa} {y}$.
Substituting $\rescost {\xb} y$ 
and $\rescost {\xa} y$ for the two bracketed terms, respectively,
we have
\begin{equation}\label{eqn:bound}
\rescost {\xb} y - \rescost {\xa} y ~\ge~ c(\xa\wedge y) - c(\xb).
\end{equation}
Note that the left-hand side is
the decrease in the residual cost for $y$ in this iteration,
which we want to show is at least $\beta$.
The right-hand side is the cost increase 
when $x$ is raised to $\xa\wedge y$
(i.e., each $x_j$ for $j\in\vars(S)$ is raised to $\min(\xa_j, y_j)$).

To complete the proof for the case $y\ge \xb$,
we show that the right-hand side is at least $\beta$.

Recall that if $y$ is feasible, then there must be 
at least one $x_k$ with $k\in\vars(S)$ and $x_k < y_k$.
\begin{description}
\item[Subcase 1 --]
When also $\xa_k < y_k$ for some $k\in\vars(S)$.   
The intuition in this case is that raising $x$ to $\xa \wedge y$
raises $x_k$ to $\xa_k$, which alone costs $\beta$
(by \ref{alg:submodular}).  Formally,
let $z$ be $\xb$ with just $\xb_k$ raised to $\xa_k$.
Then:\smallskip \\
\begin{tabular}{@{}r@{~~}r@{~\,}c@{~\,}l@{~~}l}
(i)\lefteqn{\mbox{~~\ref{alg:submodular} chooses $\xa_k$ maximally such that $c(z) \le c(\xb) + \beta$.}\hfill}
\\
(ii) & $c(z)$&$=$&$c(\xb) + \beta$
     & because (i) holds,  $c$ is continuous, and $\xa_k < \infty$. 
\\
(iii) & $z$ &$\le$&$\xa\wedge y$
  &because $z\le \xa$ and (using $x\le y$ and $x'_k < y_k$)
 $z \le y$.
\\
(iv) & $c(z)$ &$\le$& $c(\xa\wedge y)$
      & because $c$ is non-decreasing, and (iii) holds.
\end{tabular}
\smallskip 

Substituting (ii) into (iv) gives $c(\xb) + \beta \le c(\xa\wedge y)$,
that is,  $c(\xa\wedge y) - c(\xb) \ge \beta$.

\smallskip

\item[Subcase 2 -- ]
Otherwise $\xa \ge\inS y$.  
The intuition in this case is that $\xa\wedge y =\inS y$,
so that raising $x$ to $\xa\wedge y$ is enough to bring $x$ into $S$.
And, by the assumption on $\beta$ in \ref{alg:submodular}, 
it costs at least $\beta$ to bring $x$ into $S$.

Here is the formal argument.
Let $z= \xa \wedge y$.   Then:
\smallskip

\begin{tabular}{@{}r@{~~}r@{~}c@{~}ll}
(a) & $z $&$=\inS\!\!$&$ y$
   & by the definition of $z$ and $\xa \ge\inS y$.
\\
(b) & $z$&$\in$&$ S$
 & by (a), $y\in S$, and Observation~\ref{obs:up}.
\\
(c) & $z$&$\ge$&$ \xb$
& by the definition of $z$ and $\xa \ge \xb$ and $y\ge \xb$.
\\
(d) & $\rescost {\xb} S $&$\le$&$ c(z) - c(\xb)$
& by (b), (c), and the definition of $\rescost {\xb} S$.
\\
(e) & $\beta $&$\le$&$ \rescost {\xb} S$
& by the definition of \ref{alg:submodular}.
\end{tabular}
\smallskip

By transitivity, (d) and (e) imply $\beta \le c(z) - c(\xb)$,
that is, $c(\xa\wedge y) - c(\xb) \ge \beta$.
\end{description}
For the remaining case (when $y\not\ge x$),
we show that the case $y\ge x$ implies this case.
The intuition is that if $y_j < x_j$,
then $\rescost{x}{y}$ is unchanged
by raising $y_j$ to $x_j$,
so we may as well assume $y\ge x$.
Formally,
define $\hat y = y\vee \xb\ge y$.
Then $\hat y\ge x$ and $\hat y$ is feasible.

By calculation,
~ $\rescost {\xb} y 
= c(\xb \vee y) - c(\xb) 
= c(\xb \vee (y \vee \xb)) - c(\xb) 
= \rescost {\xb} {\hat y}.
$

By calculation,
 $\rescost {\xa} y 
= c(\xa \vee y) - c(\xa) 
= c(\xa \vee (y \vee \xb)) - c(\xa) 
= \rescost {\xa} {\hat y}
$.

Thus, $\rescost {\xb} y - \rescost {\xa} y$ 
equals $\rescost {\xb} {\hat y} - \rescost {\xa} {\hat y}$,
which by the case already considered is at least $\beta$.
\end{Proof}

\section{Online Covering, Paging, and Caching}
\label{sec:online}
Recall that in online \prob{Submodular-Cost Covering},
each constraint $S\in\calC$ is revealed one at a time;
an online algorithm must raise variables in $x$ to bring $x$ into the given $S$,
without knowing the remaining constraints.
\ref{alg:simple} or \ref{alg:submodular}
can do this, so by Thm.~\ref{thm:generic}
they yield  $\degree$-competitive online algorithms.\footnote
{If the cost function is linear,
in responding to $S$
this algorithm needs to know only $S$ and the values
of variables in $S$ and their cost coefficients.
In general, the algorithm needs to know $S$,
the entire cost function,
and all variables' values.}

\begin{corollary}
\label{cor:online}
\ref{alg:simple} and \ref{alg:submodular}
give $\degree$-competitive deterministic online algorithms
for \prob{Submodular-Cost Covering}.
\end{corollary}

Using simple variants of the reduction of \prob{Weighted Caching} to online \prob{Set Cover}
from \cite{Bansal07A-Primal-Dual},
Corollary~\ref{cor:online} naturally generalizes 
a number of known results for \prob{Paging}, \prob{Weighted Caching}, \prob{File Caching},
\prob{Connection Caching}, etc.\ as described in the introduction.
To illustrate such a reduction, consider the following \prob{Connection Caching} problem.
A request sequence $r$ is given online.
Each request $r_t$ is a subset of the nodes in a network.
In response to each request $r_t$,
a connection is activated (if not already activated) between all nodes in $r_t$.
Then, if any node in $r_t$ has more than $k$ active connections,
some of the connections (other than $r_t$) must be deactivated 
(paying $\cost(r_s)$ to deactivate connection $r_s$)
to leave each node with at most $k$ active connections.

Reduce this problem to online \prob{Set Cover} as follows.
Let variable $x_t$ indicate whether connection $r_t$ is
closed before the next request to $r_t$ after time $t$,
so the total cost is $\sum_t \cost(r_t) x_t$.
For each node $u$ and each time $t$,
for any $(k+1)$-subset $Q\subseteq \{r_s ~|~ s\le t; u\in r_s\}$,
at least one connection $r_s\in Q-\{r_t\}$
(where $s$ is the time of the most recent request to $r_s$)
must have been deactivated, so the following constraint\footnote
{We assume the last request must stay cached.
If not, don't subtract $r_t$ from $Q$ in each constraint.
The competitive ratio is $k+1$.}
is met:
$\max_{r_s\in Q-\{r_t\}} x_s \ge 1$.

This is an instance of online \prob{Set Cover},
with a set for each time $t$ (corresponding to $x_t$)
and an element for each triple $(u,t,Q)$ (corresponding to the constraint for that triple
as described above).

\ref{alg:simple} (via Corollary~\ref{cor:online})
gives the following $k$-competitive algorithm.
In response to a connection request $r_t$,
the connection is activated and $x_t$ is set to 0.
Then, as long as any node, say $u$, has $k+1$ active connections,
the current $x$ violates the constraint for the triple $(u,Q,t)$,
where $Q$ contains $u$'s active connections.
Node $u$ implements an iteration of \ref{alg:simple} for the violated constraint:
for all connections $r_s\in Q-\{r_t\}$,
it simultaneously raises $x_s$ at rate $1/\cost(r_s)$,
until some $x_s$ reaches $1$.
Node $u$ then arbitrarily deactivates 
connections $r_s\in Q$ with $x_s = 1$
so that at most $k$ of $u$'s connections remain active.

For a more involved example with a detailed analysis,
see Section~\ref{sec:upgradable}.

\paragraph{Remark: On $k/(k-h+1)$-competitiveness.}
The classic competitive ratio of $k/(k-h+1)$ 
(versus $\opt$ with cache size $h\le k$)
can be reproduced in the above settings as follows.  
For any set $Q$ as described above,
$\opt$ must meet the stronger constraint
$\sum_{r_s\in Q-\{r_t\}} \lfloor x_s\rfloor \ge k-h+1$.
In this scenario,
the proof of Lemma~\ref{lemma:simple}
extends to show a ratio of $k/(k-h+1)$
(use that the variables are in $[0,1]$,
so there are at least $k-h+1$
variables $x_j$ such that $x_j < y_j$,
so $\rescost x y$ decreases at rate at least $k-h+1$).

\subsection{Covering constraint generality; upgradable online problems}
\label{sec:upgradable}
Recall that the covering constraints in \prob{Submodular-Cost Covering}
need not be convex, only closed upwards.  This makes them relatively
powerful.
The main purpose of this section is to illustrate this power,
first by describing a simple example modeling file-segment requests in the {\tt http:}~protocol,
then by using it to model {\em upgradable} online caching problems.

\paragraph{Http file segment requests.}
The {\tt http:}~protocol allows retrieval of segments of files.
To model this, consider each file $f$ as a group of arbitrary segments (e.g.\ bytes or pages).
Let $x_t$ be the {\em number of segments} of file $r_t$ evicted before its next request.
Let $c(r_t)$ be the cost to retrieve a single segment of file $r_t$,
so the total cost is $\sum_t x_t \,c(r_t)$.
Then (for example), if the cache can hold at most $k$ segments total,
model this with constraints of the form (for a given subset $Q$)
$\sum_{s\in Q}\max\{0, \size(r_s) - \lfloor x_s \rfloor\} \le k$
(where $\size(r_s)$ is the total number of segments in $r_s$).

Running \ref{alg:simple} on an online request sequence gives the following online algorithm.
At time $t$, respond to the file request $r_t$ as follows.
Bring all segments of $r_t$ into the cache.
Until the current set of segments in cache becomes cacheable,
increase $x_s$ for each file with a segment in cache (other than $r_t$) at rate $1/c(r_s)$.
Meanwhile, whenever $\lfloor \min(x_s,\size(r_s)) \rfloor$ increases for some $x_s$, 
evict segment $\lfloor x_s\rfloor$ of $r_s$.
Continue until the segments remaining in cache are cacheable.

The competitive ratio will be the maximum number of files in the cache.
(In contrast, the obvious approach of modeling each segment 
as a separate cacheable item will give competitive ratio equal to
the maximum number of individual segments ever in cache.)

\newcommand{\conf}{\gamma}

\paragraph{Upgradable caching.}
The main point of this section is to illustrate the wide variety of
online caching problems that can be reduced to online covering,
and then solved via algorithms such as \ref{alg:simple}.

An \prob{Upgradable Caching} instance is specified by a maximum cache size $k$,
a number $d$ of hardware components,
the eviction-cost function $\cost(\cdots)$,
and, for each time step $t$ (revealed in an online fashion)
a request $r_t$ and a cacheability predicate, $\cacheable_t(\cdots)$.
As the online algorithm proceeds, 
it chooses not only how to evict items,
but also how to upgrade the hardware configuration.
The hardware configuration is modeled abstractly by a vector $\conf\in \Rp^d$,
where $\conf_i$ is the cost spent so far on upgrading the $i$th hardware component.
Upgrading the hardware configuration is modeled by increasing the $\conf_i$'s,
which (via $\cost(\cdots)$ and $\cacheable(\cdots)$),
can decrease item eviction costs and increase the power of the cache.

In response to each request, 
if the requested item $r_t$ is not in cache, it is brought in.
The algorithm can then increase any of the $\conf_i$'s arbitrarily
(increasing a given $\conf_i$ models spending to upgrade the $i$th hardware component).
The algorithm must then evict items (other than $r_t$) from cache
until the set $Q$ of items remaining in cache is {\em cacheable},
that is, it satisfies the given predicate $\cacheable_t(Q,\conf)$.
The cost to evict any given item $r_s$ is $\cost(r_s,\conf)$ for $\conf$
at the time of eviction.

The eviction-cost function $\cost(\cdots)$ and
each predicate $\cacheable_t(\cdots)$ must meet the following monotonicity restrictions.
The eviction-cost function $\cost(r_s, \lambda)$ must be monotone non-increasing
with respect to each $\lambda_i$.
(Intuitively, upgrading the hardware can only decrease eviction costs.)
The predicate $\cacheable_t(Q, \lambda)$ must be monotone 
with respect to $Q$ and each $\lambda_i$.
That is, increasing any single $\lambda_i$ cannot cause
the value of the predicate to switch from true to false.
(Intuitively, upgrading the hardware can only increase the power of the cache.)
Also, if a set $Q$ is cacheable (for a given $t$ and $\conf$)
then so is every subset $Q'\subseteq Q$.
Finally, for simplicity of presentation, we assume that
every cacheable set has cardinality $k$ or less.

The cost of a solution is the total paid to evict items,
plus the final hardware configuration cost, $\sum_{i=1}^d \conf_i$.
The competitive ratio is defined with respect to the minimum
cost of any sequence of evictions that meets all the specified cacheability 
constraints.\footnote
{This definition assumes that the request sequence and cacheability
requirements are independent of the responses of the algorithm.
In practice, even for standard paging, this assumption might not hold.
For example, a fault incurred by one process may cause another process's
requests to come earlier.  In this case, the optimal offline strategy
would choose responses that take into account the effects on
inputs at subsequent times (possibly leading to a lower cost).
Modeling this accurately seems difficult.}
Note that the offline solution may as well fix
the optimal hardware configuration at the start, before the first request,
as this maximizes subsequent cacheability and minimizes subsequent eviction costs.

Standard \prob{File Caching}
is the special case when
$\cacheable_t(Q,\conf)$ is the predicate
``$\sum_{r_s\in Q} \size(r_s) \le k$''
and $\cost(r_s,\conf)$ depends only on $r_s$; that is, $d=0$.
Using \prob{Upgradable caching} with $d=0$,
one could model independent use of the cache by some interfering process:
the cacheability predicate could be changed to
$\cacheable_t(Q) \equiv$ ``$\sum_{r_s\in Q} \size(r_s) \le k_t$'',
where each $k_t$ is at most $k$ but otherwise depends arbitrarily on $t$.
Or, using \prob{Upgradable caching} with $d=1$,
one could also model a cache that starts with size 1,
with upgrades to larger sizes (up to a maximum of $k$)
available for purchase at any time.
Or, also with $d=1$, one could model that upgrades of the network
(decreasing the eviction costs of arbitrary items arbitrarily)
are available for purchase at any time.
One can also model fairly arbitrary restrictions on cacheability:
for example (for illustration), one could require that, at odd times $t$, 
two specified files cannot both be in cache together, etc.

Next we describe how to reduce \prob{Upgradable Caching} 
to online \prob{Submodular-Cost Covering}
with $\degree = d+k$,
giving (via \ref{alg:simple}) a $(d+k)$-competitive online algorithm
for \prob{Upgradable Caching}.
The resulting algorithm is a natural generalization of existing algorithms.

\begin{theorem}[\bf upgradable caching]
\prob{Upgradable Caching} has a $(d+k)$-competitive online algorithm,
where $d$ is the number of upgradable components
and $k$ is the maximum number of files ever held in cache.
\end{theorem}
\begin{Proof}
Given an arbitrary \prob{Upgradable Caching} instance
with $T$ requests,
define a \prob{Submodular-Cost Covering} instance $(c,\calC)$ 
over $\Rp^{d+T}$ as follows.

The variables are as follows.
For $i=1,\ldots,d$, variable $\conf_i$ is the amount invested in component $i$.
For $t=1,\ldots,T$,
variable $x_t$ is the cost (if any) incurred for evicting the $t$th requested item $r_t$ 
at any time before its next request.  
Thus, a solution is a pair $(\conf,x)\in\Rp^d\times\Rp^T$.
The cost function is $c(\conf,x) = \sum_{i=1}^d \conf_i+\sum_{t=1}^T x_t$.

At any time $t$, let $A(t)$ denote the set of times of {\em active} requests,
the times of the most recent requests to each item:
\[A(t) = \{ s ~|~  s \le t, ~ (\forall s' \le t)~ r_{s'} = r_s \rightarrow s' \le s\}.\]
In what follows, in the context of the current request $r_t$ at time $t$,
we abuse notation by identifying each time $s\in A(t)$ with its requested item $r_s$.
(This gives a bijection between $A(t)$ and the requested items.)

For any given subset $Q\subseteq A(t)$ of the currently active items,
and any hardware configuration $\conf$,
either the set $Q$ is cacheable 
or at least one item $s\in Q-\{t\}$ must be evicted by time $t$.
In short, any feasible solution $(\conf,x)$ must satisfy the predicate
\[
\cacheable_t\big(Q,\conf)
\mbox{~~\bf or~~}
\exists s\in Q-\{t\} \mbox{ such that } x_s \ge \cost(r_s,\conf).
\]
For a given $t$ and $Q$,
let $S_t(Q)$ denote the set of solutions $(\conf,x)$ satisfying the above predicate.
The set $S_t(Q)$ is closed upwards
(by the restrictions on $\cacheable$ and $\cost$) and so is a valid covering constraint.

The online algorithm adapts \ref{alg:simple}, as follows.
It initializes $\conf=x=\mathbf 0$.
After request $r_t$, the algorithm keeps in cache the set of active
items whose eviction costs have not been paid, which we denote $C$:
\[ C = C_t(\conf,x) = \{t\}\cup \{s\in A(t) ~|~ x_s < \cost(r_s,\conf)\}.\]
To respond to request $r_t$,
as long as the cached set $C$ is not legally cacheable
(i.e., $\cacheable_t(C,\conf)$ is false),
the corresponding constraint, $S_t(C)$ is violated,
and the algorithm performs an iteration of \ref{alg:simple} 
for that constraint.
By inspection, this constraint depends on the following variables:
every $\lambda_i$, and each $x_s$ where $r_s$ is cached and $s\ne t$
(that is, $s\in C-\{t\}$).
Thus, the algorithm increases these variables at unit rate,
until either {\em (a)} $x_s \ge \cost(r_s,\conf)$ for some cached $r_s$
and/or {\em (b)} $\cacheable_t(C,\conf)$ becomes true 
(due to items leaving $C$ and/or increases in $\conf$).
When case {\em (a)} happens, the algorithm evicts that $r_s$ to maintain the invariant 
that the cached set $C$, then continues with that the new constraint for the new $C$.
When case {\em (b)} happens, the currently cached set is legally cacheable,
and the algorithms is done responding to request $t$,

This completes the description of the algorithm.
For the analysis, we define the constraint collection $\calC$ in the
underlying \prob{Submodular Covering} instance $(c,\calC)$ to contain
{\em just those constraints $S_t(C)$ for which the algorithm, 
given the request sequence, does steps}.   
When the algorithm does a step at time $t$,
the cached set $C$ contains 
only $t$ and items that stayed in cache (and where collectively cacheable) 
after the previous request.
Since at most $k$ items stayed in cache,
by inspection, the underlying constraint $S_t(C)$
depends on at most $d+k$ variables in $(\conf,x)$.
Thus, the degree $\degree$ of $(c,\calC)$ is at most $d+k$.

For the \prob{Submodular-Cost Covering} instance $(c,\calC)$,
let $(\conf^*,x^*)$  and $(\conf',x')$, respectively, be the solutions 
corresponding to $\opt$ and generated by the algorithm, respectively.
For the original upgradable caching instance (distinct from $(c,\calC)$),
let $\opt$ and $\calA$ denote the costs of, respectively, the optimal solution 
and the algorithm's solution.

Then $\calA \le c(\conf',x')$ because the algorithm paid
at most $x'_s$ to evict each evicted item $r_s$.
(We use here that $x_s \ge \cost(r_s,\conf)$ at the time of eviction,
and $x_s$ does not decrease after that;
note that $x_s$ may exceed $\cost(r_s,\conf)$
because some items with positive $x'_s$ might not be evicted.)
The approximation guarantee for \ref{alg:simple} (Lemma~\ref{lemma:simple})
ensures $c(\conf',x') \le \degree\, c(\conf^*, x^*)$.

By transitivity
\(
\calA
~\le~ c(\conf',x')
~\le~ \degree\, c(\conf^*,x^*)
~=~ \degree\, \opt.
\)
\end{Proof}

\paragraph{Flexibility in tuning the algorithm.}
In practice, it is well known that a competitive ratio much lower than $k$ 
is desirable and usually achievable for paging.  Also, for file caching
(where items have sizes), carefully tuned variants
of \prob{Landlord} (a.k.a.~\prob{Greedy-Dual Size})
outperform the original algorithms \cite{dilley99enhancement}.
In this context, it is worth noting that the above algorithm can be adjusted,
or tuned, in various ways while keeping its competitive ratio,

First, there is flexibility in how the algorithm handles ``free'' requests
--- requests to items that are already in the cache.
When the algorithm is responding to request $r_t$,
let ${t'}$ be the most recent time that item was requested
but was not in the cache at the time of the request.
Let $F(t) = \{s ~|~ t' < s < t, r_s = r_t\}$ denote the times
of these recent free requests to the item.
Worst-case sequences have no free requests,
and, although each free request $r_s$ costs nothing,
the analysis in the proof above charges $x_s$ for it anyway.

The algorithm in the proof stops the step for the current constraint $S_t(C)$
and removes an item $s$ from the cache $C$ when some $x_s$ reaches $\cost(r_s, \conf)$.
Modify the algorithm to stop the step (and remove $s$ from $C$) sooner,
specifically, when $x_s$ reaches $\cost(r_s, \conf) - \sum_{s'\in F(s)} x_{s'}$ for some $s\in C$
(effectively reducing the eviction cost of $r_s$ by $\sum_{s'\in F(s)} x_{s'}$.
The modified algorithm is still a specialization of \ref{alg:simple}.
Although the resulting solution $(x,\conf)$ may be infeasible,
the approximation guarantee still applies, 
in that $(x,\conf)$ has cost at most $\degree\,\opt$.
The online solution $\calA$ {\em is} feasible though,
and has cost equal to the cost of $(x,\conf)$,
and is thus $\degree$-competitive.

In the above description, each free request is used to reduce the effective cost
of a later request to the same item.  Whereas the unmodified algorithm generalizes
LRU, the modified algorithm generalizes FIFO.  

Even more generally,
the sum over the free requests $r_s$ of $x_s$ can be {\em arbitrarily} distributed
over the non-free requests to reduce their effective costs (leading to earlier eviction).
Essentially the same analysis still shows $k$-competitiveness.

\smallskip
There is a second, 
independent source of flexibility --- the rates at which the variables are increased in each step.
As it specializes for \prob{File Caching}, the algorithm in the proof 
raises each $x_s$ at unit rate until $x_s$ reaches $\cost(r_s)$.
This raises the total cost $c(x,\conf)$ at rate $\sum_{s\in C-\{t\}} 1 \le k$,
while (in the analysis of \ref{alg:simple} the residual cost of $\opt$ decreases at rate at least 1,
implying a competitive ratio of $k$.
In contrast, \prob{Landlord} (effectively) raises each $x_s$ at rate $\size(r_s)$ 
until $x_s$ reaches $\cost(r_s)$.
This raises $c(x,\conf)$ more rapidly, at rate $\sum_{s\in C-\{t\}} \size(r_s)$,
but this sum is also at most $k$
(since all summed items fitted in the cache before $r_t$ was brought in).
This implies the (known) competitive ratio of $k$ for \prob{Landlord}.
Generally, for items of size larger than 1, 
the algorithm could raise $x_s$ at any rate in $[1,\size(r_s)]$.
The more general algorithm still has competitive ratio at most $k$.

Analogous adjustments can be made in other applications of \ref{alg:simple}.
For some applications, adjusting the variables' relative rates of increase 
can lead to stronger theoretical bounds.

\section{Stateless Online Algorithm 
and Randomized Generalization of \ref{alg:submodular}}
\label{sec:randomized} \label{sec:random}
This section describes two randomized algorithms
for \prob{Submodular-Cost Covering}:
\ref{alg:stateless} ---
a {\em stateless} 
$\degree$-competitive online algorithm,
and an algorithm that generalizes both that and \ref{alg:submodular}.
For simplicity, in this section we assume each $U_j$ has finite cardinality.
(The algorithms can be generalized in various ways to arbitrary closed $U_j$,
but the presentation becomes more technical.\footnote
{
Here is one of many ways to modify \ref{alg:stateless} to handle arbitrary closed $U_j$'s.
In each step, take $\beta$ small enough
so that for each $j\in \vars(S)$, 
either $U_j$ contains the entire interval $[x_j,x_j+\beta]$, 
or $U_j$ contains just $x_j$ from that interval.
For the latter type of $j$, take $\beta_j$ and $\xhat_j$ as described in \ref{alg:stateless}.
For the former type of $j$, take $\beta_j=\beta$ and take $\xhat_j$ to be the smallest value
such that increasing $x_j$ to $\xhat_j$ would increase $c(x)$ by $\beta$.
Then proceed as above.
(Taking $\beta$ infinitesmally small gives the following process.
For each $j\in \vars(S)$ simultaneously, $x_j$ increases continuously
at rate inversely proportional to its contribution to the cost,
if it is possible to do so while maintaining $x_j\in U_j$,
and otherwise $x_j$ increases to its next allowed value randomly
according to a Poisson process
whose intensity is inversely proportional to the resulting expected increase in the cost.)
})

\ref{alg:stateless} generalizes 
the \prob{Harmonic} $k$-server algorithm as it specializes for \prob{Paging} and \prob{Caching}
\cite{Raghavan94Memory},
and Pitt's \prob{weighted vertex cover} algorithm \cite{Bar-Yehuda00One-for-the-price}.

\begin{definition}[\bf {\em stateless} online algorithm]
An online algorithm for a (non-canonical) \prob{Submodular-Cost Covering} 
instance $(c,U,\calC)$ is {\em stateless}
provided the only state it maintains is the current solution $x$,
in which each $x_j$ is assigned only values in $U_j$.
\end{definition}
Although \ref{alg:simple} and \ref{alg:submodular} 
maintain only the current partial solution $x\in\Rp^n$,
for problems with variable-domain restrictions
$x_j$ may take values outside $U_j$.
So these algorithms are not stateless.\footnote
{The online solution is not $x$,
but rather $x'\le x$ defined from $x$ by $x'_j = \max\{\alpha\in U_j ~|~ \alpha \le x_j\}$
or something similar, 
so the algorithms maintain state other than the current online solution $x'$.
For example, for paging problems, the algorithms maintain $x_t\in [0,1]$ as they proceed,
where a requested item $r_s$ is currently evicted only once $x_s = 1$.
To be stateless,
they should maintain each $x_t\in\{0,1\}$,
where $x_s=0$ iff page $r_s$ is still in the cache.}

\begin{figure}[t]
\begin{alg}
\Ahead{\bf Stateless algorithm for Submodular-cost Covering}
\Alabel{alg:stateless}
\Ain{cost $c$, finite domains $U$, constraints $\calC$}
\A Initialize $x_j \leftarrow \min U_j$ for each $j$.
\A In response to each given constraint $S$, repeat the following until $x\in S$:
\algbeg
\A For each $j\in\vars(S)$:
\algbeg
\A If $x_j < \max U_j$:
\algbeg
\A Let $\xhat_j = \min\{\alpha \in U_j ~|~ \alpha > x_j\}$ be the next largest value in $U_j$.
\A Let $\beta_j$ be the increase in $c(x)$ that would result from raising $x_j$ to $\xhat_j$.
\algend
\A Else:
\algbeg
\A Let $\xhat_j = x_j$ and $\beta_j = \infty$.
\A If $(\forall j\in\vars(S))~\beta_j=\infty$: Return ``infeasible''.
\algend
\algend
\Along{Increase $x_j$ to $\xhat_j$ for all $j\in J$,
  where $J$
  is any random subset of $\vars(S)$
  such that, for some $\beta\ge 0$,
  for each $j\in\vars(S)$,
  $\Pr[j\in J] = \beta/\beta_j$.
  Above interpret $0/0$ as $1$.
 (Note that there are many ways to choose $J$ with the necessary property.)
}
\end{alg}
\end{figure}
The stateless algorithm initializes each $x_j$ to $\min U_j$.
Given any constraint $S$, it repeats the following until $S$ is satisfied:
it chooses a random subset $J\subseteq \vars(S)$,
then increases each $x_j$ for $j\in J$ to its next allowed value,
$\min\{\alpha \in U_j ~|~ \alpha > x_j\}$.
The subset $J$ can be any random subset such that,
for some $\beta\ge 0$,
for each $j\in\vars(S)$,
$\Pr[j\in J]$ equals $\beta/\beta_j$,
where $\beta_j$ is the increase in $c(x)$ that would result from increasing $x_j$.
\smallskip

For example, 
one could take $J=\{r\}$ where
$r$ is chosen so that $\Pr[r = j] \propto 1/\beta_j$.
Or take any $\beta\le \min_j \beta_j$,
then, independently for each $j\in\vars(S)$,
take $j$ in $J$ with probability $\beta/\beta_j$.
Or, choose $\tau\in[0,1]$ uniformly,
then take $J= \{ j \giv \beta/\beta_j \ge \tau\}$.
In the case that each $U_j = \{0,1\}$ and $c$ is linear,
one natural special case of the algorithm is to repeat the following as long as there is 
some unsatisfied constraint $S$:

{\em 
Choose a single $k\in \{ j ~|~ j\in \vars(S), x_j=0\}$ at random,
so that $\Pr[ k=j ] \propto 1/c_j$.
Set $x_k=1$.
}

\begin{theorem}[\bf correctness of stateless \ref{alg:stateless}]\label{thm:stateless}
For online \prob{Submodular-Cost Covering} with finite variable domains,
\ref{alg:stateless} is stateless.
If the step sizes are chosen so the number of iterations has finite expectation
(e.g.~taking $\beta=\Omega(\min_j \beta_j)$),
then it is $\degree$-competitive (in expectation).
\end{theorem}

\newcommand{\xtt}[1]{x^{#1}}
\newcommand{\xt}{{\xtt t}}
\newcommand{\xz}{{\xtt 0}}
\newcommand{\xT}{{\xtt T}}
\newcommand{\upj}[1]{{\,\uparrow_{\!_J}^{#1}\,}}

\begin{Proof}
By inspection the algorithm maintains each $x_j\in U_j$.
It remains to prove $\degree$-competitiveness.

Consider any iteration of the repeat loop.
Let $\xb$ and $\xa$, respectively, denote $x$ before and after the iteration.
Let $\beta$ and $\beta_j$ be as in the algorithm.

First we observe that iteration increases the cost of algorithm's solution $x$
by at most $\beta\degree$ in expectation:

\smallskip
\noindent
{\bf Claim 1:}
{\em Cost $c(x)$ increases by at most
$\sum_{j\in\vars(S)} (\beta/\beta_j) \beta_j = \beta|\vars(S)| \le \beta\degree$
in expectation.}

\smallskip
\noindent
The claim follows easily by direct calculation and the submodularity of $c$.

\medskip
Inequality (\ref{eqn:bound}) from the proof of Thm.~\ref{thm:generic} still holds:
$\rescost {\xb} y - \rescost {\xa} y ~\ge~ c(\xa\wedge y) - c(\xb)$,
so the next claim implies
that the residual cost of any feasible $y\ge x$
decreases by at least $\beta$ in expectation:
\smallskip

\noindent
{\bf Claim 2:} 
For any feasible $y\ge \xb$,  $E_J[c(\xa\wedge y) - c(\xb) ~|~ \xb] \ge \beta$.    

\smallskip
\noindent
{\em Proof of claim.}
By Observation~\ref{obs:up}, there is a $k\in\vars(S)$ with $y_k > x_k$.
Since $y_k \in U_k$, the algorithm's choice of $\xhat_k$ ensures $y_k \ge \xhat_k$.
Let $z$ be obtained from $\xb$ by raising just $\xb_k$ to $\xhat_k$.
With probability $\beta/\beta_k$, the subroutine raises $x_k$ to $\xhat_k \le y_k$,
in which case $c(\xa \wedge y) - c(\xb) \ge c(z) - c(\xb) = \beta_k$.
This implies $E_J[c(\xa \wedge y) - c(\xb)~|~\xb] \ge (\beta/\beta_k) \beta_k = \beta$,
proving Claim 2.

\medskip

Thus, for $y\ge x$, in each iteration, the residual cost of $y$ decreases by at least $\beta$ in expectation: $E_J[\rescost {\xb} y  - \rescost {\xa} y ~|~ \xb] \ge \beta$.
By the argument at the end of the proof of Thm.~\ref{thm:generic},
this implies the same for {\em all} feasible $y$ (even if $y\not\ge \xb$).

In sum, the iteration increases the cost of $x$ by at most $\degree\beta$ in expectation,
while decreasing the residual cost of any feasible $y$ by at least $\beta$ in expectation.
By standard probabilistic arguments, this implies that the expected final cost of $x$
is at most $\degree$ times the initial residual cost of $y$ (which equals the cost of $y$).

Formally, $c(\xt) + \degree\rescost {\xt} y$ is a super-martingale,
where random variable~$\xt$ denotes $x$ after $t$ iterations.

Let random variable~$T$ be the number of iterations.
Using, respectively, $\rescost {\xT} y \ge 0$,
a standard optional stopping theorem,
and $\rescost {\xz} y = c(y) - c(\xz)$ (because $\xz \le y$),
the expected final cost $E[c(\xT)]$ is at most

\smallskip\noindent
$
E[c(\xT) + \degree\,\rescost{\xT}y]
~\le~
E[c(\xz) + \degree\,\rescost{\xz}y]
~=~ c(\xz) + \degree\,\big(c(y) - c(\xz)\big)
~\le~\degree\,c(y).
$
\end{Proof}

\paragraph{Most general randomized algorithm.}
\ref{alg:submodular} raises the variables continuously,
whereas \ref{alg:stateless} steps each variable $x_j$ through the successive values in $U_j$.
For some instances, both of these choices can lead to slow running times.
Next is an algorithm that generalizes both of these algorithms.
The basic algorithm is simple, but the condition on $\beta$ is more subtle.
The analysis is a straightforward technical generalization of the previous analyses.

The algorithm has more flexibility in increasing variables.
This may be important in distributed or parallel applications,
where the flexibility allows implementing the algorithm
so that it is guaranteed to make rapid (probabilistic) progress.
(The flexibility may also be useful for dealing with limited-precision arithmetic.)

The algorithm is \ref{alg:submodular},
modified to call subroutine $\rstep_c(x,S)$ (\ref{alg:rstep}, below)
instead of $\step_c(x,S)$ 
to augment $x$ in each iteration.

\medskip
\begin{alg}
\Ahead{{\bf Subroutine} $\rstep_c$}
\Alabel{alg:rstep}
\Ain{current solution $x\in\Rpi^n$, unsatisfied constraint $S\in\calC)$}
\A Fix an arbitrary probability $p_j \in [0,1]$ for each $j \in \vars(S)$.
\\\comment{above, taking each $p_j=1$ gives \ref{alg:submodular}}
\A Choose a step size $\beta \ge 0$
where $\beta$ is at most expression (\ref{eqn:random})
in  Thm.~\ref{thm:random}.
\A For $j$ with $p_j>0$, let $\xhat_j$ be maximum~such that raising $x_j$ to $\xhat_j$
would raise $c(x)$ by at most $\beta/p_j$.
\A Choose a random subset\footnote
{As in \ref{alg:stateless}, the events ``$j\in J$'' for $j\in\vars(S)$ can be dependent.
See the last line of \ref{alg:stateless}.}
$J\subseteq\vars(S)$ s.~t.~ $\Pr[j\in J] = p_j$ for $j\in \vars(S)$.
\label{line:J}
\A For $j\in J$, let $x_j\leftarrow \xhat_j$. \vspace*{-0.5em}
\end{alg}

The step-size requirement is a bit more complicated.

\begin{theorem}[\bf correctness of randomized algorithm]\label{thm:random}
For \prob{Submodular-Cost Covering}
suppose, in each iteration of the randomized algorithm 
for a constraint $S\in\calC$ and $x\not\in S$,
the step size $\beta\ge 0$ is at most
\begin{equation}\label{eqn:random}
\min~\big\{~ E_J\big[c(x \upj{y})-c(x)\big] 
~~:~~ y\ge x;~~y\in S~\big\},
\end{equation}
where $x\upj{y}$ is a random vector obtained by
choosing a random subset $J$
from the same distribution used in Line~\ref{line:J} of \rstep
and then raising $x_j$ to $y_j$ for $j\in J$.
Suppose also that the expected number of iterations is finite.
Then the algorithm returns a $\degree$-approximate solution in expectation.
\end{theorem}
Note that if $p=\mathbf 1$, then (\ref{eqn:random})
simplifies to $\rescost x S$.
If $c$ is linear,  (\ref{eqn:random})
simplifies to $\rescostp x S$ where $c'_j = p_j c_j$.
\begin{Proof}
The proof mirrors the proof of Thm.~\ref{thm:stateless}.

Fix any iteration.
Let $\xb$ and $\xa$, respectively, denote $x$ before and after the iteration.
Let $p$, $\beta$, $\xhat$, and $J$ be as in $\rstep$.

\smallskip

\noindent{\bf Claim 1.}  
{\em The expected increase in $c(x)$ is

$E_J[c(\xa)-c(\xb)|\xb] ~\le~ \sum_{j\in\vars(S)} p_j \beta/p_j ~=~ \beta|\vars(S)| ~\le~ \beta\degree$.}

\smallskip\noindent
The claim follows easily by calculation and the submodularity of $c$.

\medskip
\noindent
Inequality (\ref{eqn:bound}) from the proof of Thm.~\ref{thm:generic} still holds:
$\rescost {\xb} y - \rescost {\xa} y ~\ge~ c(\xa\wedge y) - c(\xb)$,
so the next claim implies
that the residual cost of any feasible $y\ge x$
decreases by at least $\beta$ in expectation:

\medskip\noindent{\bf Claim 2.}  {\em For any feasible $y\ge \xb$,~
$E_J[c(\xa\wedge y) - c(\xb)~|~ \xb] \ge \beta$.   }

\smallskip
\noindent{\em Proof of claim:}
The structure of the proof is similar to the corresponding part of the proof of Thm.~\ref{thm:generic}.

Recall that if $y$ is feasible, then there must be 
at least one $x_k$ with $k\in\vars(S)$ and $x_k < y_k$.

\begin{description}
\item[Subcase 1 --]
When also there is an $\xhat_k < y_k$ for $k\in\vars(S)$ with $p_k>0$.   

In case of the event $k\in J$, raising $x$ to $\xa\wedge y$ raises $x_k$ to $\xhat_k$,
which alone (by \ref{alg:rstep}) costs $\beta/p_k$. 

Thus, the expected cost to raise $x$ to $\xa\wedge y$ 
is at least $\Pr[k \in J]\, \beta/p_k = \beta$.
\smallskip

\item[Subcase 2 --]
Otherwise, $\xhat_j \ge y_j$ for all $j\in J$ (for all possible $J$).

In this case, $\xa\wedge y \ge \xb \upj{y}$ in all outcomes.

Thus, the expected cost to increase $\xb$ to $\xa\wedge y$
is at least the expected cost to increase $\xb$ to $\xb \upj{y}$.

By the assumption in the theorem, this is at least $\beta$.
This proves Claim 2.
\end{description}
Claims 1 and 2 imply $\degree$-approximation via the argument in
the final paragraphs of the proof of Thm.~\ref{thm:stateless}.
\end{Proof}

\section{Relation to local-ratio method}
\label{sec:local}

The local-ratio method has most commonly been applied
to problems with variables taking values in $\{0,1\}$
and with linear objective function $c\cdot x$
(see
\cite{Bar-Yehuda85A-local-ratio,Bar-Yehuda00One-for-the-price,Bar-Yehuda05On-the-equivalence,Bar-Yehuda04Local};
for one exception, see \cite{Bar-Yehuda01Efficient}).
For example, \cite{Bar-Yehuda05On-the-equivalence}
shows a form of equivalence between the primal-dual method
and the local-ratio method, but that result only considers problems
with solution space $\{0,1\}^n$ (i.e., 0/1-variables).
Also, the standard intuitive interpretation of local-ratio
--- that the algorithm reduces the coefficients in the cost vector $c$ ---
works only for 0/1-variables.

Here we need to generalize to more general solution spaces.
To begin, we first describe a typical local-ratio algorithm 
for a problem with variables over $\{0,1\}$ (we use \prob{CIP-01}).
After that, we describe one way to extend the approach to more general variable domains.
With that extension in place, we then recast Thm.~\ref{thm:generic} 
(the approximation ratio for \ref{alg:submodular}) as a local-ratio analysis.

\paragraph{Local-ratio for $\{0,1\}$ variable domains.}
Given a (non-canonical) \prob{Linear-Cost Covering} instance $(c,U,\calC)$
where each $U_j = \{0,1\}$, the standard local-ratio approach gives the following
$\degree$-approximation algorithm:
\begin{quote}
{\em Initialize vector $\ell = c$.  
Let ``the cost of $x$ under $\ell$'' be $\sum_j \ell_j x_j$.
Let $\hat x(\ell)$ be the maximal $x\in\{0,1\}^n$ that has zero cost under $\ell$
(i.e., $\hat x_j(\ell)=1$ if $\ell_j = 0$).
As long as $\hat x(\ell)$ fails to meet some constraint $S\in\calC$, repeat the following:
Until $\hat x(\ell) \in S$,
simultaneously for all $j\in\vars(S)$ with $\ell_j>0$, decrease $\ell_j$ at unit rate.
Finally, return $\hat x(\ell)$.}
\end{quote}
The algorithm has approximation ratio $\degree=\max_S |\vars(S)|$
by the following argument.  Fix the solution $\xalg$ returned by the algorithm.
An iteration for a constraint $S$ 
decreases $\ell_j \xalg_j$ for each $j\in\vars(S)$ at rate $\xalg_j \le 1$,
so it decreases $\ell \cdot \xalg$ at rate at most $\degree$.
On the other hand, in any feasible solution $x^*$,
as long as the variables $x_j$ for $j\in S$ are being decreased,
at least one $j\in\vars(S)$ with $\ell_j>0$ has $x^*_j=1$
(otherwise $\hat x(\ell)$ would be in $S$).
Thus the iteration decreases $\ell \cdot x^*$ at rate at least 1.
From this it follows that $c\cdot \xalg \le \degree c\cdot x^*$
(details are left as an exercise).

This local-ratio algorithm is the same as \ref{alg:simple} for the case $U=\{0,1\}^n$
(and linear cost).
To see why, observe that the modified cost vector $\ell$ in the local-ratio algorithm
is implicitly keeping track of the residual problem for $x$ in \ref{alg:simple}.
When the local-ratio algorithm reduces a cost $\ell_j$ at unit rate,
for the same $j$, \ref{alg:simple} increases $x_j$ at rate $1/c_j$.
This maintains the mutual invariant $(\forall j)~~\ell_j = c_j(1-x_j)$
--- that is, $\ell_j$ is the cost to raise $x_j$ the rest of the way to 1.
Thus, as they proceed together,
{\em the \prob{CIP-01} instance $(\ell, \calC)$
defined by the current (lowered) costs $\ell$ is exactly
the residual problem $(\resc x, \calC)$ for the current $x$ in \ref{alg:simple}.}
To confirm this, note that the cost of any $y$ in the residual problem for $x$
is $\rescost x y = \sum_j c_j\max(y_j-x_j, 0) = \sum_{j : y_j = 1} c_j (1-x_j)$,
whereas in the local-ratio algorithm the cost for $y$ under $\ell$ is
$\sum_{j:y_j=1} \ell_j$, and by the mutual invariant above these are equal.

So, at least for linear-cost covering problems with $\{0,1\}$-variable domains,
we can interpret local-ratio via residual costs, and vice versa.   
On the other hand, residual costs extend naturally to more general domains.
Is it possible to likewise extend the local-ratio cost-reduction approach?
Simply reducing some costs $\ell_j$ until some $\ell_j=0$ does not work
--- $\ell_j\xalg_j$ may decrease at rate faster than $1$,
and when $\ell_j$ reaches 0, it is not clear which value $x_j$ should take in $U_j$.

\paragraph{Local ratio for more general domains.}
One way to extend local-ratio to more general variable domains is as follows.
Consider any (non-canonical) instance $(c,U,\calC)$ where $c$ is linear.
Assume for simplicity that each variable domain $U_j$
is the same: $U_j = \{0,1,\ldots,u\}$ for some $u$ independent of $j$,
and that all costs $c_j$ are non-zero.
For each variable $x_j$, instead of maintaining a single reduced cost $\ell_j$,
the algorithm will maintain a {\em vector} $\ell_j\in\Rp^u$ of reduced costs.
Intuitively, $\ell_{jk}$ represents the cost to increase $x_j$ from $k-1$ to $k$.
(We are {\em almost} just reducing the general case to the 0/1 case 
by replacing each variable $x_j$ by multiple copies,
but that alone doesn't quite work, as it increases $\degree$ by a factor of $u$.)
Define the cost of any $x\in \{0,1,\ldots,u\}^n$ under the current $\ell$ to be
\(\sum_j \sum_{k=1}^{x_j} \ell_{jk}\).
As a function of the reduced costs $\ell$,
define $\hat x(\ell)$ to be the maximal zero-cost solution,
i.e. $\hat x_j(\ell) = \max\{ k ~|~ \sum_{i=1}^k \ell_{ji} = 0\}$.

The local-ratio algorithm initializes each $\ell_{jk} = c_j$,
so that the cost of any $x$ under $\ell$ 
equals the original cost of $x$ (under $c$).
The algorithm then repeats the following until $\hat x(\ell)$ satisfies all constraints.
\begin{quote}
1. Choose any constraint $S$ that $\hat x(\ell)$ does not meet.  
Until $\hat x(\ell)\in S$, do:
\\2. ~~Just until an $\ell_{jk}$ reaches zero,
for all $j\in \vars(S)$ with $\hat x_j(\ell) < u$
\\3. ~~~simultaneously,
lower $\ell_{jk_j}$ at unit rate, where $k_j= \hat x_j(\ell) + 1$.
\end{quote}
Finally the algorithm returns $\hat x(\ell)$ 
(the maximal $x$ with zero cost under the final $\ell$).

One can show that this algorithm is a $\degree$-approximation algorithm
(for $\degree$ w.r.t.~the original \prob{CIP-UB} instance)
by the following argument.
Fix $\xalg$ and $x^*$ to be, respectively, the algorithm's final solution and an optimal solution.
In an iteration for a constraint $S$, as $\ell$ changes,
the cost of $\xalg$ under $\ell$ decreases at rate at most $\degree$,
while the cost of $x^*$ under $\ell$ decreases at rate at least 1.
We leave the details as an exercise.

In fact, the above algorithm is equivalent to \ref{alg:simple} for \prob{CIP-UB}.
If the two algorithms are run in sync,
at any given time, the \prob{CIP-01} instance with modified cost $\ell$ 
exactly captures the residual problem for \ref{alg:simple}.

\paragraph{Local-ratio for Submodular-Cost Covering.}
The previous example illustrates the basic ideas underlying one approach
for extending local-ratio to problems with general variable domains:
decompose the cost into parts, one for each possible increment of each variable,
then, to satisfy a constraint $S$, for each variable $x_j$ with $j\in\vars(S)$,
lower just the cost for that variable's next increment.
This idea extends somewhat naturally even to infinite variable domains,
and is equivalent to the residual-cost interpretation.

Next we tackle \prob{Submodular-Cost Covering} in full generality.
%
%
We recast the proof of Thm.~\ref{thm:generic} 
(the correctness of \ref{alg:submodular}) as a local-ratio proof.
Formally, the minimum requirement for the local-ratio method 
is that the objective function can be decomposed into ``locally approximable'' objectives.
The common cost-reduction presentation of local ratio described above
gives one such decomposition,  but others have been used (e.g.~\cite{Bar-Yehuda01Efficient}).
In our setting, the following local-ratio decomposition works.
(We discuss the intuition after the lemma and proof.)
\smallskip
\begin{lemma}[\bf local-ratio lemma]\label{lemma:decompose}
Any algorithm returns a $\degree$-approximate solution $x$ provided
there exist $T\in\Zp$ and $c^t:\Rp^n\rightarrow \Rp$ (for $t=1,2,\ldots,T$) and $r:\Rp^n\rightarrow \Rp$ such that
\smallskip


(a)~ for any $y$, $c(y) = \sum_{t=1}^T c^t(y) ~\,+\,r(y)$,

(b)~ for all $t$, and any $y$ and feasible $x^*$,  $c^t(y) \le c^t(x^*)\degree$,

(c)~ the algorithm returns $x$ such that $r(x) = 0$.
\end{lemma}
\begin{Proof}
Properties (a)-(c) state that the cost function can be decomposed into parts,
where, for each part $c^t()$, {\em any} solution $y$ is $\degree$-approximate,
and, for the remaining part $r()$, the solution $x$ returned by the algorithm has cost zero.
Since $x$ is $\degree$-approximate w.r.t.~each $c^t()$, 
and $x$ has cost zero for the remaining part, $x$ is $\degree$-approximate overall.
Formally, let $x^*$ be an optimal solution.
By properties (a) and (c), (b), then (a), respectively,
\smallskip

\hfill\(\SPREAD
c(x) 
\Eq
\sum_{t=1}^T c^t(x)
\Le
\sum_{t=1}^T c^t(x^*) \degree ~+~ r(x^*)\degree
\Eq
c(x^*)\degree.
\)\hfill
\end{Proof}
In local-ratio as usually presented, 
the local-ratio algorithm determines the cost decomposition as it proceeds.
The only state maintained by the algorithm after iteration $t$
is the ``remaining cost'' function $\ell^t$,
defined by $\ell^t(y) = c(y) - \sum_{s\le t} c^s(y)$.
In iteration $t$, the algorithm 
determines some portion $c^t$ of $\ell^{t-1}$
satisfying Property (b) in the lemma
and removes it from the cost.
(This is the key step in designing the algorithm.)
The algorithm stops when it has removed enough of the cost
so that there is a feasible solution $\xalg$ with
zero remaining cost ($\ell^T(\xalg) = 0$), then returns that $\xalg$
(taking $r = \ell^T$ for Property (c) in the lemma).
By the lemma, this $\xalg$ is a $\degree$-approximate solution.

For a concrete example, consider
the local-ratio algorithm for the linear-cost, 0/1-variable case 
described at the start of this section.  
Let $T$ be the number of iterations.
For $t=0,1,\ldots,T$, let $\ell^t$ be the modified cost vector at the end of iteration $t$
(so $\ell^0$ is the original cost vector).
Define $c^t(y)=(\ell^t-\ell^{t-1})\cdot y$
to be the decrease in the cost of $y$ due to the change in $\ell$ in iteration $t$.  
Define $r(y)=\ell^T\cdot y$ to be the modified cost vector at termination
(so the returned solution $x=\hat x(\ell^T)$ has $r(x) = 0$).
It is easy to see that property (a) and (c) hold.
To see that property (b) holds, recall that in iteration $t$
the algorithm reduces all $\ell_j$ for $j\in\vars(S)$ with $\ell_j>0$,
simultaneously and continuously at unit rate.
It raises each $x_j$ to 1 when $\ell_j$ reaches 0.  It stops once $x\in S$.
At most $\degree$ of the $\ell_j$'s are being lowered at any time,
so the rate of decrease in $\ell\cdot y$ for any $y\in\{0,1\}^n$ is at most $\degree$.
But for any $x^* \in S$, the rate of decrease in $\ell\cdot x^*$ is at least $1$,
because at least one $j\in\vars(S)$ has $x^*_j = 1$ and $\ell_j > 0$
(otherwise $x$ would be in $S$).

Next we describe how to generate such a decomposition of the cost $c$
corresponding to a run of \ref{alg:submodular}
on an arbitrary \prob{Submodular-Cost Covering} instance $(c,\calC)$.
This gives an alternate proof of Thm.~\ref{thm:generic}.
The proof uses the previously described idea for extending local ratio 
to more general domains.
Beyond that, it is slightly more complicated than the argument in the previous
paragraph for two reasons: it handles submodular costs,
and, more subtly, in an iteration for a constraint $S$,
\ref{alg:submodular} can increase variables more than enough to satisfy $S$
(of course this is handled already in the previous analysis of \ref{alg:submodular},
which we leverage below).

\begin{lemma}[\bf correctness of \ref{alg:submodular} via local-ratio]\label{lemma:local}
\ref{alg:submodular}, run on any instance $(c,\calC)$ of  \prob{Submodular-Cost Covering},
implicitly generates a cost decomposition $\{c^t\}$ and $r$
as described in Lemma~\ref{lemma:decompose}.
Thus, \ref{alg:submodular} gives a $\degree$-approximation.
\end{lemma}
\begin{Proof}[Proof sketch.]
Assume without loss of generality that $c(\mathbf 0) = 0$.
(Otherwise use cost function $c'(x) = c(x) - c(\mathbf 0)$.
Then $c'(x)$ is still non-negative and non-decreasing,
and, since $\Delta\ge 1$, the approximation ratio for $c'$ implies it for $c$.)

Let $x^t$ denote \ref{alg:submodular}'s vector $x$ after $t$ iterations.
Let $T$ be the number of iterations.

Recall that $\resc{x^t}$ is the cost in the residual problem
$(\resc{x^t},\calC)$ for $x$ after iteration $t$:
$\rescost {x^t} y = c(x^t\vee y) - c(x^t)$.

Define $c^t$ so that the ``remaining cost'' function $\ell^t$ (as discussed before the lemma)
equals the objective $\resc{x^t}$ in the residual problem for $x^t$.
Specifically, take $c^t(y) = \rescost {x^{t-1}} y - \rescost {x^t} y$.  
Also define $r(y)=\rescost{x^T} y$.

These $c^t$ and $r$ have properties (a-c) from Lemma~\ref{lemma:decompose}.

Properties (a) and (c) follow by direct calculation.
To show (b), fix any $y$.
Then
$c^t(y) = c(x^t) - c(x^{t-1}) 
+ c(x^{t-1}\vee y) - c(x^{t}\vee y)
\le c(x^t) - c(x^{t-1})$.
So $c^t(y)$ is at most the increase in the cost $c(x)$ of $x$ during iteration $t$.
In the proof of Thm.~\ref{thm:generic}, 
this increase in $c(x)$ in iteration $t$
is shown to be at most $\degree\beta$.
Also, for any feasible $x^*$, 
the cost $\rescost x {x^*}$ for $x^*$ in the residual problem for $x$
is shown to reduce by at least $\beta$.
But the reduction in $\rescost x {x^*}$ is exactly $c^t(x^*)$.
Thus, $c^t(y) \le \degree\beta \le \degree c^t(x^*)$, proving Property (b).
\end{Proof}

Each $c^t$ in the proof captures the part of the cost $c$ lying ``between'' $x^{t-1}$ and $x^t$.
For example, if $c$ is linear, then $c^t(y) = \sum_j c_j \big|[0,y_j]\cap[x_j^{t-1},x_j^t]\big|$.
The choice of $x^t$ in the algorithm guarantees property (b) in the lemma.

\section{Relation to primal-dual method; local valid inequalities}
\label{sec:relation}
\label{sec:primaldual}
Next we discuss how \ref{alg:simple} can be reinterpreted as a primal-dual algorithm.

It is folklore that local-ratio and primal-dual algorithms are ``equivalent'';
for example \cite{Bar-Yehuda05On-the-equivalence}
shows a formal equivalence between the primal-dual method
and the local-ratio method.   
But that result only applies to problems with solution space $\{0,1\}^n$ (i.e., 0/1-variables),
and the underlying arguments do not seem to extend directly to this more general setting.

Next we present two linear-program relaxations for
\prob{Linear-Cost Covering},
then use the second one to reprove Lemma~\ref{lemma:simple} 
(that \ref{alg:simple} is a $\degree$-approximation algorithm
for \prob{Linear-Cost Covering})
using the primal-dual method.

Fix any \prob{Linear-Cost Covering} instance $(c,\calC)$ in canonical form.

To simplify the presentation, assume 
at least one optimal solution to $(c,\calC)$ 
is finite (i.e., in $\Rp^n$).

For any $S\in\calC$, let $\overline S$ denote the complement of $S$ in $\Rpi^n$.
Let $\overline S^*$ denote the closure of $\overline S$ under limit.

By Observation~\ref{obs:up},
if $x$ is feasible,
then, for any $S\in\calC$ and $y\in\overline S$,
$x$ meets the {\em non-domination} constraint $x \not <\inS y$
(that is, $x_j\ge y_j$ for some $j\in\vars(S)$).
By a limit argument,\footnote
{If $x\in S$ and $y\in\overline S^*$, then $y$ is the limit of some sequence $\{y^t\}$ of points in $\overline S$.   
Each $y^t$ has $x^t_{j(t)} \ge y^t_{j(t)}$ for some $j(t)\in\vars(S)$.  
Since $|\vars(S)|$ is finite, for some $j\in\vars(S)$, the infinite subsequence 
$\{y^t ~|~ j(t)=j\}$ also has $y$ as a limit point.  Then $y_j$ is the limit of the $y^t_j$'s
in this subsequence, each of which is at most $x_j$, so $y_j$ is at most $x_j$.}
the same is true if $y\in\overline S^*$.
In sum, if $x$ is feasible, then $x$ meets the non-domination constraint 
for every $(S,y)$ where  $S\in\calC$ and $y\in\overline S^*$.
For finite $x$, the converse is also true:
\begin{observation}\label{obs:nondom}
If $x\in\Rp^n$ meets the non-domination constraint
for every $S\in\calC$ and $y\in\overline S^*$,
then $x$ is feasible for $(c,\calC)$.
\end{observation}
\begin{Proof}
Assume $x$ is not feasible.  Fix an $S\in\calC$ with $x\not\in S$.
Define $y(\eps)$ by $y_j(\eps) = x_j + \eps$ so $\lim_{\eps\rightarrow 0} y = x \not\in S$.
Since $S$ is closed under limit, $y(\eps')\not\in S$ for some $\eps'>0$.
Since $x$ is finite, $x_j < y_j(\eps')$ for each $j\in\vars(S)$.
Thus, $x <\inS y(\eps')$
(i.e., $x$ fails to meet the non-domination constraint for $(S, y(\eps'))$).
\end{Proof}

\paragraph{First relaxation.}
The non-domination constraints suggest this relaxation of $(c,\calC$):
\[
\minimize~~ c\cdot x \mbox{~~~subject to~~~}
(\forall S\in\calC, y\in \overline S^*)~~~
\sum_{j\in \vars(S)} {x_j}/{y_j} ~\ge~ 1.
\]
Let $(c,\calR^1)$ denote this \prob{Linear-Cost Covering} instance.
Call it {\em Relaxation 1}.
\begin{observation}\label{obs:relaxation1}
Fix any $x\in\Rp^n$ that is feasible for $(c,\calR^1)$.

Then $\degree\, x$ is feasible for $(c,\calC)$.
\end{observation}
\begin{Proof}
Fix any $S\in\calC$ and $y\in\overline S^*$.

Then $\sum_{j\in\vars(S)} x_j/y_j \ge 1$.
Thus, $\max_{j\in\vars(S)} x_j/y_j \ge 1/|\vars(S)|$.

Thus,  $\max_{j\in\vars(S)} \degree x_j/y_j \ge 1$.

That is, $\degree x$ meets the non-domination constraint for (any) $(S,y)$.

By Observation~\ref{obs:nondom}, $\degree x$ is feasible for $(c,\calC$).
\end{Proof}

\begin{corollary}[\bf relaxation gap for first relaxation]\label{cor:firstgap}
The relaxation gap\footnote
{The relaxation gap is the maximum, 
over all instances $(c,\calC)$ of \prob{\footnotesize Linear-Cost Covering},
of the ratio [optimal cost for $(c,\calC)$] /
[optimal cost for its relaxation $(c,\calR^1)$].}
for $(c,\calR^1)$ is at most $\degree$.
\end{corollary}
\begin{Proof}
Let $x$ be a finite optimal solution for $(c,\calR^1)$.
By Obs.~\ref{obs:relaxation1}, $\degree\, x$ is feasible for $(c,\calC)$,
and has cost $c\cdot (\degree x) = \degree (c\cdot x)$.
Thus, the optimal cost for $(c,\calC)$ is at most $\degree$
times the optimal cost for $(c,\calR^1)$.
\end{Proof}

\noindent
Incidentally, $(c,\calR^1)$ gives an ellipsoid-based
\prob{Linear-Cost Covering}
$\degree$-approximation algorithm.\footnote
{Briefly, run the ellipsoid method to solve $(c,\calR^1)$
using a separation oracle that, given $x$, 
checks whether $\degree\, x \in S$ for all $S\in\calC$,
and, if not, returns an inequality that $x$ violates for $\calR^1$
(from the proof of Observation~\ref{obs:relaxation1}).
Either the oracle finds, for some $x$, that $\degree\, x\in S$ for all $S$,
in which case $\xa = \degree\, x$  
is a $\degree$-approximate solution for $(c,\calC)$,
or the oracle returns to the ellipsoid method a sequence of violated inequalities
that, collectively, prove that $(c,\calR^1)$ (and thus $(c,\calC)$) is infeasible.
}

\paragraph{Linear-Cost Covering reduces to Set Cover.}  
From the \prob{Linear-Cost Covering} instance $(c,\calC)$,
construct an equivalent (infinite) \prob{Set Cover} instance $(c',(E,\calF))$ as follows.
Recall the non-domination constraints:
$x\not<\inS y$ for each $S\in\calC$ and $y\in\overline S^*$.
Such a constraint is met if,
for some $j\in\vars(S)$, $x_j$ is assigned a value $r\ge y_j$.
Introduce an element $e=(S,y)$ into the element set $E$ for each pair $(S,y)$ 
associated with such a constraint.
For each $j\in[n]$ and $r\in\Rp$, introduce a set $s(j,r)$ into the set family $\calF$,
such that set $s(j,r)$ contains element $(S,y)$ if assigning $x_j=r$
would ensure $x\not <\inS y$
(i.e., would satisfy the non-domination constraint for $(S,y)$).
That is, $s(j,r) = \{ (S,y) ~|~ j\in\vars(S),~r \ge y_j\}$.
Take the cost of set $s(j,r)$ to be $c'_{jr} = rc_j$
(equal to the cost of assigning $x_j=r$).

\begin{observation}[\bf reduction to Set Cover]\label{obs:setcover}
The \prob{Linear-Cost Covering} instance $(c,\calC)$ 
is equivalent to the above \prob{Set Cover} instance $(c',(E,\calF))$.
By ``equivalent'' we mean that each feasible solution $x$ to $(c,\calC)$
corresponds to a set cover $X$ for $(E,\calF)$ 
(where $s(j,r)\in X$ iff $x_j = r$) and, conversely,
each set cover $X$ for $(E,\calF)$
corresponds to a feasible solution $x$ to $(c,\calC)$ 
(where $x_j = \sum_{r : s(j,r) \in X} r$).
Each correspondence preserves cost.
\end{observation}

The observation is a consequence of Observation~\ref{obs:nondom}.

Note that above reduction increases $\degree$.

\paragraph{Second relaxation, via Set Cover.}
Relaxation 2 is the standard LP relaxation of \prob{Set Cover},
applied to the equivalent \prob{Set Cover} instance $(c',(E,\calF))$ above,
with a variable $X_{jr}$ for each set $s(j,r)\in \calF$:
\[
\minimize \sum_{j,r} r c_j X_{jr}
\mbox{~~subject to~~}
~(\forall S\in\calC, y\in\overline S^*)
~~
 \sum_{j\in \vars(S)} \sum_{r \ge y_j} X_{jr} \, \ge\, 1.
\]
(There is a technicality in the definition above ---
the index $r$ of the inner sum ranges over $[y_j,\infty)$.
Should one sum, or integrate, over $r$?
Either can be appropriate --- the problem
and its dual will be well-defined
and weak duality will hold either way.
Here we restrict attention to solutions $X$ 
with finite support, so we sum.
The same issue arises in the dual below.)

We denote the above relaxation $(c',\calR^2)$.
By Observation~\ref{obs:setcover},
any feasible solution $x$ to $(c,\calC)$ gives a feasible solution
to $(c,\calR^2)$ of the same cost (via $X_{jr} = 1$ iff $r=x_j$
and $X_{jr}=0$) otherwise).
Incidentally, any feasible solution $X$ to $(c',\calR^2)$ 
also gives a solution $x$ to ($c,\calR^1)$
of the same cost, via $x_j = \sum_r r X_{jr}$.
That is, Relaxation 1 is a relaxation of Relaxation 2.
The converse is not generally true.\footnote
{The instance $(c,\calC)$ defined by
 $\min\{x_1 + x_2 ~|~ x\in\Rp^2;~x_1+x_2 \ge 1\}$
has optimum cost 1.
In its first relaxation $(c,\calR^1)$,
$x_1 = x_2 = 1/4$ with cost 1/2 is feasible.
But one can show (via duality) that $(c',\calR^2)$
has optimal cost at least 1.
}

\paragraph{Dual of Set-Cover relaxation.}  
The linear-programming dual of Relaxation 2
is the standard \prob{Set Cover} dual:
fractional packing of elements under (capacitated) sets.
We use a variable $z_e$ for each element $e$:
\[\maximize \sum_{e\in E} z_e \mbox{~~subject to~~}
(\forall~s(j,r)\in \calF)~~
\sum_{e \in s(j,r)} z_e \le r c_j
.\]
Recall $E=\{(S,y) ~|~ S\in\calC, y\in\overline S^*\}$;
$s(j,r) = \{ (S,y)\in E ~|~ j\in\vars(S),~r \ge y_j\}$.

\smallskip
We now describe the primal-dual interpretation of \ref{alg:simple}.
\begin{lemma}[\bf primal-dual analysis of \ref{alg:simple}] 
\ref{alg:simple} can be augmented to compute, along with the solution
$x$ to $(c,\calC)$, a solution $z$ to the dual of Relaxation 2
such that $c\cdot x$ is at most $\degree$ times the cost of $z$.
Thus, \ref{alg:simple} is a $\degree$-approximation algorithm.
\end{lemma}
\begin{Proof}
Initialize $z=\mathbf 0$.
Consider an iteration of \ref{alg:simple} for some constraint $S'$.
Let $\xb$ and $\xa$, respectively, be the solution $x$ before and after the iteration.
Fix element $e' = (S',\xa)$.
Augment \ref{alg:simple} to raise\footnote
{In fact this dual variable must be 0 before this, because $\xa_j>\xb_j$ for some $j$,
so this dual variable has not been raised before.}
the dual variable $z_{e'}$ by $\beta$.
This increases the dual cost by $\beta$.
Since the iteration increases the cost of $x$ by at most $\beta\degree$,
the iteration maintains the invariant that the cost of $x$ is at most $\degree$ 
times the dual cost.

To finish, we show the iteration maintains dual feasibility.
For any element $e=(S,y) \in E$, let $S(e)$ denote $S$.
Increasing the dual variable $z_{e'}$ by $\beta$ 
maintains the following invariant:
\[\textstyle
\mbox{for all } j\in[n], ~~~x_j c_j = \sum_{e : j\in \vars(S(e))} z_e.
\] 
The invariant is maintained because
$z_{e'}$ occurs in the sum iff $j\in\vars(S(e'))=\vars(S')$,
and each $x_j$ is increased (by $\beta/c_j$)
iff $j\in\vars(S')$,
so the iteration increases both sides of the equation equally.

Now consider any dual constraint that contains the raised variable $z_{e'}$.
Fix the pair $(j,r)$ defining the dual constraint.
That $e'\in s(j,r)$ implies $j\in \vars(S')$ and $\xa_j \le r$.
Each dual variable $z_{e}$ that occurs in this dual constraint has $j\in\vars(S(e))$.
But, by the invariant, at the end of the iteration,
the sum of {\em all} dual variables $z_{e}$ with $j\in\vars(S(e))$
equals $\xa_j c_j$.
Since $\xa_j \le r$, this sum is at most $r c_j$.
Thus, the dual constraint remains feasible at the end of the iteration.
\end{Proof}

\subsection{Valid local inequalities; the ``price of locality''}
\label{sec:locality}
Here is one general way of characterizing the analyses in this paper in terms of valid inequalities.
Note that each of the valid inequalities that is used in Relaxation 1
from Section~\ref{sec:primaldual} can be obtained by considering 
some single constraint ``$x\in S$'' in isolation, 
and adding valid inequalities for just that constraint.    
Call such a valid inequality ``local''.
This raises the following question: What if we were to add {\em all} local valid inequalities (ones that can be obtained by looking at each $S$ in isolation)?   
What can we say about the relaxation gap of the resulting polytope?

Formally, fix any \prob{Submodular-Cost Covering} instance
\(\min \{c(x) ~|~ x \in S \mbox{ for all } S \in \calC\}\).   
Consider the ``local'' relaxation $(c,\calL)$ obtained as follows.
For each constraint $S\in\calC$, 
let $\conv(S)$ denote the convex closure of $S$.
Then let $\calL = \{ \conv(S) ~|~ S\in\calC \}$.
Equivalently, for each $S\in\calC$,
let $\calL_S$ contain all of the linear inequalities on variables in $\vars(S)$
that are valid for $S$,
then let $\calL = \bigcup_{S\in\calC} \calL_S$.
For \prob{Linear-Cost Covering}, 
Relaxation 1 above is a relaxation of $(c,\calL)$,
so Corollary~\ref{cor:firstgap} implies that the gap is at most $\degree$.
It is not hard to find examples\footnote
{
Here is an example in $\R^2$.
For $v\in \R^2$, let $|v|$ denote the 1-norm $\sum_i |v_i|$.
For each $v\in\Rp^2$ such that $|v|= 1$,
define constraint set $S_v = \{ x\in \Rp^2  : (\exists j) x_j \ge v_j\}$.
Consider the covering problem $\min\{|x| : (\forall v) x\in S_v \}$.

Each constraint $x\in S_v$ excludes points dominated by $v$,
so the intersection of all $S_v$'s is $\{x \in \Rp^2 : |x| \ge 1\}$.
On the other hand, since $S_v$ contains the points $(v_1,0)$ and $(0,v_2)$,
$\conv(S_v)$ must contain $x = v_2 (v_1, 0) + v_1(0,v_1) = (v_1v_2, v_1v_2)$,
where $v_1v_2 \le (1/2)^2 = 1/4$.
Thus, each $\conv(S_v)$ contains $x=(1/4,1/4)$, with $|x|=1/2$.
Thus, the relaxation gap of $(c,\calL)$ for this instance is at least 2.

Another example with $\degree=2$, this time in $\Rp^n$.
Consider the sets $S_{ij} = \{x\in\Rp^n : \max(x_i,x_j) \ge 1\}$.
Consider the covering problem $\min\{|x|: (\forall i,j) x\in S_{ij} \}$.
Each point $x \in \bigcap_{ij} S_{ij}$ has $|x^*| \ge (n-1)/n$,
but $x=(1/2,1/2,1/2,\ldots,1/2)$ is in each $\conv(S)$, and $|x|=n/2$,
so the relaxation gap of $(c,\calL)$ is at least 2.
}
showing that the gap is at least $\degree$.

Of course, if we add {\em all} (not just local) valid inequalities for 
the feasible region $\bigcap_{S\in\calC} S$, 
then every extreme point of the resulting feasible region
is feasible for $(c,\calC)$, so the relaxation gap would be 1.

\section{Fast Implementations for Special Cases of Submodular-Cost Covering}
\label{sec:implementation}


This section has a linear-time implementation of \ref{alg:submodular}
for \prob{Facility Location} (and thus also for \prob{Set Cover} and \prob{Vertex Cover}),
a nearly linear-time implementation
for \prob{CMIP-UB},
and an $O(N\widehat\degree\log\degree)$-time implementation for two-stage probabilistic \prob{CMIP-UB}.  (Here $N$ is the number of non-zeroes in the constraint matrix and $\widehat\degree$ is the maximum, over all variables $x_j$,
of the number of constraints that constrain that variable.)
The section also introduces a two-stage probabilistic version of \prob{Submodular Covering},
and shows that it reduces to ordinary \prob{Submodular Covering}.

For \prob{Facility Location}, $\degree$ is the maximum number of facilities 
that might serve any given customer.
For \prob{Set Cover}, $\degree$ is the maximum set size.
For \prob{Vertex Cover}, $\degree = 2$.

\subsection{Linear-time implementations for Facility Location, Set Cover, and Vertex Cover}
\label{sec:facility_location}
The standard integer linear program for \prob{Facility Location}
is not a covering linear program due to constraints of the form ``$x_{ij} \le y_j$''.
Also, the standard reduction of \prob{Facility Location}
to \prob{Set Cover} increases $\Delta$ exponentially.
For these reasons, we formulate \prob{Facility Location} directly
as the following special case of \prob{Submodular-Cost Covering}, taking advantage of submodular cost:
\begin{quote}\em
minimize
~~$\sum_{j} f_j \max_{i} x_{ij}
~+~
\sum_{ij} d_{ij} x_{ij}$

~~subject to
(for each customer $i$)
~~$\max_{j\in N(i)} x_{ij} ~\ge~ 1$.
\end{quote}
Above $j\in N(i)$ if customer $i$ can use facility $j$.
($N(i) = \vars(S_i)$ where $S_i$ is the constraint above for customer $i$.)
\begin{theorem}[\bf linear-time implementations]
For (non-metric) \prob{Facility Location},
\prob{Set Cover}, and \prob{Vertex Cover},
the greedy $\degree$-approximation algorithm (\ref{alg:submodular}) 
has a linear-time implementation.  
\end{theorem}
\begin{Proof}
The implementation is as follows.  

\smallskip
1. Start with all $x_{ij}=0$.  Then, for each customer $i$, in any order, do the following:

2. ~~~Let $\beta =\min_{j\in N(i)} [d_{ij} + f_j(1 - \max_{i'} x_{i'j})]$
\\\hspace*{4em}(the minimum cost to raise $x_{ij}$ to 1 for any $j\in N(i)$).

3. ~~~For each $j\in N(i)$, raise $x_{ij}$ by 
$\min[ \beta / d_{ij}, (\beta + f_j \max_{i'} x_{i'j})/(d_{ij} + f_j)]$

4. Assign each customer $i$ to any facility $j(i)$ with $x_{ij(i)}= 1$.

5. Open the facilities that have customers.

\smallskip

\noindent
Line 3 raises the $x_{ij}$'s just enough to increase the cost by $\beta$ per raised $x_{ij}$
and to increase $\max_{j\in N(i)} x_{ij}$ to $1$.

By maintaining, for each facility $j$, $\max_{i} x_{ij}$,
the implementation can be done in linear time, $O(\sum_i |N(i)|)$.

\prob{Set Cover} is the special case when $d_{ij}=0$;
\prob{Vertex Cover} is the further special case $\degree=2$.
\end{Proof}

\subsection{Nearly linear-time implementation for CMIP-UB}
\label{sec:restriction}
\label{sec:CMIP}

This section describes a nearly linear-time implementation of \ref{alg:submodular}
for \prob{Covering Mixed Integer Linear Programs} with upper bounds on the variables (\prob{CMIP-UB}), that is, problems of the form
\[\min\big\{c\cdot x ~\big|~ x\in\Rp^n;~ Ax \ge B;~ x\le u;~~(\forall j\in I)~x_j\in \Z \big\},\]
where $c\in\Rp^n$, $A\in\Rp^{m\times n}$ and $B\in\Rp^n$ have no negative entries.
The set $I$ contains the indices of the variables that are restricted to take integer values,
while $u\in\Rpi^n$ gives the upper bounds on the variables.
$\degree$ is the maximum number of non-zeroes in any row of $A$.
We prove the following theorem:
\begin{theorem}[\bf implementation for CMIP-UB]\label{thm:CMIP}
For \prob{CMIP-UB},
\ref{alg:submodular} can be implemented to return a $\degree$-approximation 
in $O(N\log \degree)$ time,
where $N$ is the total number of non-zeroes in the constraint matrix.
\end{theorem}
\begin{Proof} Fix any \prob{CMIP-UB} instance as described above.
For each constraint $A_ix\ge B_i$ (each row of $A$), do the following.
For presentation (to avoid writing the subscript $i$),
rewrite the constraint as $a\cdot x \ge b$ (where $a=A_i$ and $b=B_i$).
Then bring the constraint into canonical form, as follows.
Assume for simplicity of presentation
that integer-valued variables in $S$ come before the other variables
(that is, $I\cap\vars(S) = \{1,2,\ldots,\ell\}$ for some $\ell$).
Assume for later in the proof
that these $\ell$ variables are ordered 
so that $a_1 \ge a_2 \ge \cdots \ge a_\ell$.
(These assumptions are without loss of generality.)
Now incorporate the variable-domain restrictions ($x\le u$ and $(\forall j\in I)~x_j\in\Z$)
into the constraint by rewriting it as follows:

\smallskip
$\displaystyle\SPREAD
 \sum_{j=1}^\ell a_j\lfloor \min(x_j,u_{j}) \rfloor 
+ \sum_{j>\ell} a_j \min(x_j,u_{j}) ~\Ge~ b
$.
{\hfill (canonical constraint $S$ for $A_ix\ge B_i$)}
\\

\noindent

\smallskip
Let $\calC$ be the collection of such canonical constraints,
one for each original covering constraint $A_i x\ge B_i$.

\smallskip
\noindent{\bf Intuition.}
The algorithm focuses on a single unsatisfied $S\in \calC$,
repeating an iteration of \ref{alg:submodular}
(raising the variables $x_j$ for $j\in\vars(S)$)
until $S$ is satisfied.  It then moves on to another unsatisfied $S$, 
and so on, until all constraints are satisfied.
While working with a particular constraint $S$,
it increases each $x_j$ for $j\in\vars(S)$ by $\beta/c_j$ for some $\beta$.
We must choose $\beta \le \rescost x S$
(the optimal cost to augment $x$ to satisfy $S$),
thus each step requires some lower bound on $\rescost x S$.
But the steps must also be large enough to satisfy $S$ quickly.

For intuition, consider first the case when $S$ has no variable upper bounds
(each $u_j=\infty$) and no floors.
In this case, the optimal augmentation of $x$ to satisfy $S$
simply raises the single most cost-effective variable $x_j$
(minimizing $a_j/c_j$) to satisfy $S$, so $\rescost x S$ is
easy to calculate exactly and taking $\beta=\rescost x S$ satisfies $S$ in one iteration.

Next consider the case when $S$ has some variable upper bounds (finite $u_j$).
In this case, we take $\beta$ to be the minimum cost to {\em either}
satisfy $S$ {\em or} bring some variable to its upper bound 
(we call this {\em saturating} the variable).
This $\beta$ is easy to calculate, and will satisfy $S$ after at most $\vars(S)$ iterations
(as each variable can be saturated at most once).

Finally, consider the case when $S$ also has floors.
This complicates the picture considerably.
The basic idea is to relax (remove) the floors,
satisfy the relaxed constraint as described above,
and then reintroduce the floors one by one.
We reintroduce a floor only once the constraint {\em without} that floor is already satisfied.
This ensures that {\em the constraint with the floor will be satisfied if
the term with the floor increases even once}.
(If the term for a floored variable $x_j$ increases, we say $x_j$ is {\em bumped}.)
We also reintroduce the floors in a particular order --- in order of decreasing $a_j$.
This ensures that introducing one floor (which lowers the value of the left-hand side)
does not break the property in italics above for previously reintroduced floors.

The above approach ensures that $S$ will be satisfied in $O(\vars(S))$ iterations.
A careful but straightforward use of heaps allows all the iterations for $S$ to be done in $O(\vars(S)\log\degree)$ time.  This will imply the theorem.

\smallskip
Here are the details. To specify the implementation of \ref{alg:submodular}, 
we first specify how, in each iteration,
for a given constraint $S\in\calC$ and $x\not\in S$,
the implementation chooses the step size $\beta$.
It starts by finding a relaxation $S^h$ of $S$ (that is, $S\subseteq S^h$,
so $\rescost x {S^h} \le \rescost x S$).
Having chosen the relaxation,
the algorithm then takes $\beta$ 
to be the minimum cost needed to raise any {\em single} variable $x_j$ (with $j\in\vars(S)$) 
just enough to either satisfy the relaxation $S^h$ or to cause $x_j = u_j$.

The relaxation $S^h$ is as follows.
Remove all floors from $S$,
then add in just enough floors (from left to right), so that the resulting constraint
is unsatisfied.
Let $S^h$ be the resulting constraint, where $h$ is the number of floors added in.
Formally,
For $h=0,1,\ldots,\ell$, define
\(f^h(x) \,=\, \sum_{j=1}^h a_j \lfloor \min(x_j,u_j) \rfloor + \sum_{j > h} a_j \min(x_j,u_j)\)
to be the left-hand side of constraint $S$ above,
with only the first $h$ floors retained.
Then fix $h = \min\{h\ge 0 ~|~ f^h(x) < b\}$,
and take $S^h = \{x ~|~ f^h(x) \ge b\}$.

\smallskip

Next we show that this $\beta$ satisfies the constraint in \ref{alg:submodular}.
\begin{lemma}[\bf validity of step size]
For $S$, $x\not\in S$, and $\beta$ as described above, $\beta \in\ [0,\rescost x S]$.
\end{lemma}
\begin{Proof}
  As $S\subseteq S^h$, it suffices to prove $\beta \leq \tilde{c}_x(S^h)$.  
  Recall that a variable $x_j$ is \emph{saturated} if $x_j=u_j$.
  Focus on the unsaturated variables in $\vars(S)$.  
  We must show that if we wish to augment (increase) some variables 
  just enough to saturate a variable or bring $x$ into $S^h$, 
  then we can achieve this at minimum cost by increasing a single variable.  
  This is certainly true if we saturate a variable: 
  only that variable needs to be increased. 
  A special case of
  this is when some $c_i$ is~0---we can saturate $x_i$ at zero cost, which is
  minimum.  Therefore, consider the case where all $c_i$'s are positive
  and the variable increases bring $x$ into $S^h$.  

  Let $P$ be the set of unsaturated variables in $\{x_1,\ldots,x_h\}$, 
  and let $Q$ be the set of unsaturated variables among $\{x_j ~|~ j>h\}$.  
  Consider increasing a variable $x_j\in P$.  Until $x_j$ is bumped
  (i.e., the term $\lfloor x_j \rfloor+1$ increases because $x_j$ reaches its next higher integer),
  $f^h(x)$ remains unchanged, but the cost increases.  
  Thus, if it is optimal to increase $x_j$ at all, $x_j$ must be bumped.
  When $x_j$ is bumped, $f^h(x)$ jumps by $a_j$, 
  which (by the ordering of coefficients) is at least $a_h$, 
  which (by the choice of $h$) is sufficient to bring $x$ into $S^h$.
  Thus, if the optimal augmentation increases a variable in $P$, 
  then the only variable that it increases is that one variable, which is bumped once.

  The only remaining case is when
  the optimal augmentation of $x$ increases only variables from $Q$.  
  Let $x_k=\arg\min\{c_j/a_j ~|~ x_j\in Q\}$.  Clearly it is not advantageous
  to increase any variable in $Q$ other than $x_k$.  (Let $\delta_j\geq 0$ denote the
  amount by which we increase $x_j\in Q$.  If $\delta_j>0$ for some $j\neq k$, then
  we can set $\delta_j=0$ and instead increase $\delta_k$ by $a_j\delta_j/a_k$.
  this will leave the increase in $f^h(x)$ intact, so $x$ will still be brought
  into $S^h$, yet will not inflate the cost increase, because the cost will
  decrease by $c_j\delta_j$, but increase by $c_k a_j \delta/a_k \leq
  c_j\delta_j$, where the inequality holds by the definition of $k$.)
\end{Proof}

By the lemma and Thm.~\ref{thm:generic}, with this choice of $\beta$,
the algorithm gives a $\degree$-approximation.
It remains to bound the running time.  
\begin{lemma}[\bf iterations]\label{lemma:hitting}
For each $S\in\calC$, the algorithm does at most $2|\vars(S)|$ iterations for $S$.
\end{lemma}
\begin{Proof}
Recall that, in a given iteration,
$\beta$ is the minimum such that raising some single $x_k$ by $\beta/c_k$
(with $k\in\vars(S)$ and $x_k < u_k$) is enough to saturate $x_k$ or bring $x$ into $S^h$.
If the problem is feasible, $\beta<\infty$ so there is such an $x_k$.
Each iteration increases $x_j$ for {\em all} $j\in\vars(S)$ by $\beta/c_j$,
so must increase this $x_k$ by $\beta/c_k$.
Thus, the iteration either saturates $x_k$ or brings $x$ into $S^h$.

The number of iterations for $S$ that saturate variable is clearly at most $|\vars(S)|$.
The number of iterations for $S$ that satisfy that iteration's relaxation
(bringing $x$ into $S^h$) is also at most $|\vars(S)|$, 
because, by the choice of $h$, in the next iteration for $S$ 
the relaxation index $h$ will be at least 1 larger.
Thus, there are at most $2|\vars(S)|$ iterations for $S$ before $x\in S$.
\end{Proof}

The obvious implementation of an iteration for a given constraint $S$
runs in time $O(|\vars(S)|)$
(provided the constraint's $a_j$'s are sorted in a preprocessing step).
By the lemma, the obvious implementation thus yields total time 
$O(\sum_{S} |\vars(S)|^2) 
\le O(\sum_{S} |\vars(S)| \degree) 
= O(N \degree)$.

To complete the proof of Thm.~\ref{thm:CMIP}, we show how to use standard
heap data structures to implement the above algorithm to run in $O(N \log \degree)$ time.
The implementation considers the constraints $S\in\calC$ in any order.
For a given $S$, it repeatedly does iterations for that $S$ until $x\in S$.
As the iterations for a given $S$ proceed,
the algorithm maintains the following quantities:
\begin{description}
\item[$\bullet$] A fixed vector $x^b$, which is $x$ at the start of the first iteration for $S$,
  initialized in time $O(|\vars(S)|)$.

\item[$\bullet$] A variable $\tau$, tracking the sum of the $\beta$'s for $S$ so far (initially 0).
  Crucially, the current $x$ then satisfies $x_j = x^b_j + \tau/c_j$ for $j\in\vars(S)$.
  While processing a given $S$, we use this to represent $x$ implicitly.

  We then use the following heaps to find each {\em breakpoint} of $\tau$ ---
  each value at which a variable becomes saturated, is bumped,
  or at which $S^h$ is satisfied and the index $h$ of the current relaxation $S^h$ increases.
  We stop when $S^\ell$ (that is, $S$) is satisfied.

\item[$\bullet$] A heap containing, for each unsaturated variable $x_j$ in $\vars(S)$,
  the value $c_j (u_j-x_j^b)$ of $\tau$ at which $x_j$ would saturate.
  This value does not change until $x_j$ is saturated, at which point the value is removed 
  from the heap.

\item[$\bullet$] A heap containing, for each unsaturated integer variable $x_j$ ($j\le h$) 
  in $S^h$, the value
  of $\tau$ at which $x^j$ would next be bumped.
  This value is initially $c_j(1-(x^b_j - \lfloor x^b_j\rfloor))$.
  It changes only when $x_j$ is bumped, 
  at which point it increases by $c_j$.

\item[$\bullet$] A heap containing, for each unsaturated non-integer variable $x_j$ ($j>h$)
  in $S^j$, the ratio $c_j / a_j$.
  This value does not change.  It is removed from the heap when $x_j$ is saturated.

\item[$\bullet$] The current {\em derivative} $d$ of $f^h(x)$ with respect to $\tau$, 
which is $d = \sum_{j>h, x_j < u_j} a_j/c_j$.  
This value changes by a single term whenever a variable is saturated 
or $h$ increases.

\item[$\bullet$] The current slack $b^h = b - f^h(x)$ of $S^h$,
  updated at each breakpoint of $\tau$.
\end{description}
In each iteration, the algorithm queries the min-values of each of the three heaps.
It uses the three values to calculate the minimum value of $\tau$ at which, respectively,
a variable would become saturated,
a variable would be bumped,
or a single (non-integer) variable's increase would increase $f^h(x)$ by the slack $b^h$.
It then increases $\tau$ to the minimum of these three values.
(This corresponds to doing a step of \ref{alg:simple} with $\beta$ equal to the increase in $\tau$.)
With the change in $\tau$, it
detects each saturation, bump, and increment of $h$ that occurs,
uses the derivative to compute the increase in $f^h(x)$,
then updates the data structures accordingly.
(For example, it removes saturated variables' keys from all three heaps.)

After the algorithm has finished all iterations for a given constraint $S$,
it explicitly sets $x_j \leftarrow x_j^b + \tau/c_j$ for $j\in\vars(S)$,
discards the data structures for $S$,
and moves on to the next constraint.

The heap keys for a variable $x_j$ change (and are inserted or removed) 
only when that particularly variable is bumped, or saturated, or when $h$ increases to $j$.
Each variable is saturated at most once,
and $h$ increases at most $\ell\le \vars(S)$ times,
and thus there are at most $\vars(S)$ bumps
(as each bump increases $h$ by at least 1).
Thus, during all iterations for $S$,
the total number of breakpoints and heap key changes
is $O(\vars(S))$.  Since each heap operation takes $O(\log \degree)$ time,
the overall time is then $O(\sum_{S\in\calC} |\vars(S)|\log\degree)$
$=O(N\log\degree)$, where $N$ is the number of non-zeros in $A$.

This proves the theorem.
\end{Proof}

\subsection{Two-Stage (Probabilistic) Submodular-Cost Covering}
\label{sec:submodular}

An instance of {\em two-stage} \prob{Submodular-Cost Covering}
is a tuple $(W,p,(c,\calC))$ where
$(c,\calC)$ is an instance of \prob{Submodular-Cost Covering} over $n$ variables
(so $S\subseteq \Rpi^n$ for each $S\in\calC$),
$W:\Rpi^{|\calC|\times n}\rightarrow\Rpi$ is a non-decreasing, submodular, continuous {\em first-stage} objective function,
and, for each $S\in\calC$, the {\em activation probability} of $S$ is $p_S$.
A solution is a collection $X=[x^S]_{S\in \calC}$ 
of vectors $x^S\in\Rpi^n$, one for each constraint $S\in\calC$,
such that $x^S\in S$.
Intuitively, $x^S$ specifies how $S$ will be satisfied if $S$ is activated,
which happens with probability $p_S$.
As usual $\degree = \max_{S\in\calC} |\vars(S)|$.

The solution should minimize the cost $w(X)$ of $X$, 
as defined by the following random experiment.
Each constraint $S$ is independently {\em activated} with probability $p_S$.
This defines a \prob{Submodular-Cost Covering} instance 
$(c, \calA)$ where $\calA =\{ S\in\calC ~|~ S \mbox{ is activated}\}\subseteq\calC$,
and the solution $x^\calA$ for that instance
defined by $x^\calA_j = \max\{ x^S_j ~|~ S \in \calA\}$.
Intuitively, $x^\calA$ is the minimal vector that meets the first-stage commitment
to satisfy each activated constraint $S$ with $x^S$.
The cost $w(X)$ is then $W(X) + E_{\calA}[ c(x^\calA)]$,
the first-stage cost $W(X)$ (modeling a ``preparation'' cost)
plus the (expectation of the) second-stage cost $c(x^\calA)$
(modeling an additional cost for assembling the final
solution to the second-stage \prob{Submodular-Cost Covering} instance $(c,\calA)$).

\paragraph{Facility-Location example.}
For example, consider a \prob{Set Cover} instance $(c,\calC)$
with elements $[m]$ and sets $s(j)\subseteq [m]$ for $j\in[n]$.
That is,
\(
\minimize\, c\cdot x \,\subjectto\,
x\in\Rp^n,
~ (\forall i\in[m])~\max_{j: i\in s(j)} x_j \ge 1.
\)
\smallskip

\noindent
Extend this to a two-stage \prob{Set Cover} instance $(W,p,(c,\calC))$
where $W_{ij}\ge 0$ and each $p_i=1$.
Let $X=[x^i]_i$ be any (minimal) feasible solution to this instance.
That is, $x^i\in\{0,1\}^n$ says that element $i$ chooses
the set $s(j)$ where $x^i_j = 1$.
All constraints are activated in the second stage, so
each $x^\calA_j = \max\{ x^i_j ~|~ i\in s(j)\}$.
That is, $x^\calA_j = 1$ iff any element $i$ has chosen set $s(j)$.
The cost $w(X)$ is $\sum_{ij} W_{ij} x^i_j ~+~ \sum_j c_j \max\{ x^i_j ~|~ i\in s(j)\}$.

Note that this two-stage \prob{Set Cover} problem
exactly models \prob{Facility Location}.
The first-stage cost $W$ captures the assignment cost;
the second-stage cost $c$ captures the opening cost.

\medskip
Consider again general  two-stage \prob{Submodular-Cost Covering}.
A $\degree$-approximation algorithm for it follows immediately from the following observation:
\begin{observation}\label{obs:2stage}
Two-stage \prob{Submodular-Cost Covering} reduces to \prob{Submodular-Cost Covering} (preserving $\degree$).
\end{observation}
\begin{Proof}
Any two-stage instance $(W,p,(c,\calC))$ over $n$ variables
is equivalent to a standard instance $(w, \calC')$ over $n|\calC|$ variables
($X=[x^S]_{S\in\calC}$)
where $w(X)$ is the cost of $X$ for the two-stage instance as defined above,
and, for each $S\in\calC$, there is a corresponding constraint $x^S\in S$ on $X$ in $\calC'$.
One can easily verify that the cost $w(X)$ is submodular, non-decreasing, and continuous
because $W(X)$ and $c(x)$ are.  
\end{Proof}

Next we describe a fast implementation of
\ref{alg:submodular} for  two-stage \prob{CMIP-UB} ---
the special case of two-stage \prob{Submodular-Cost Covering}
where $W$ is linear and the pair $(c,\calC)$ form a \prob{CMIP-UB} instance.
\begin{theorem}[\bf implementation for two-stage CMIP-UB] \label{thm:2stage}
For two-stage \prob{CMIP-UB}:

(a) \ref{alg:submodular} can be implemented to return a $\degree$-approximation
in $O(N\widehat\degree\log\degree)$ time,
where
$\widehat\degree$ is the maximum number of constraints per variable
and $N$ is the input size
$\sum_{S\in\calC}|\vars(S)|$.

(b) When $p=\mathbf 1$,
the algorithm can be implemented to run in time $O(N\log \degree)$.
(The case $p=\mathbf 1$ of two-stage \prob{CMIP-UB}
generalizes \prob{CMIP-UB} and \prob{Facility Location}).
\end{theorem}
\begin{Proof}
Fix an instance $(W,p,(c,\calC))$ of two-stage \prob{CMIP-UB}.
Let $(w,\calC')$ be the equivalent instance of standard \prob{Submodular-Cost Covering} 
from Observation~\ref{obs:2stage} over variable vector $X=[x^S]_{S\in\calC}$.
Let random variable $x^\calA$ be as described in the problem definition
($x^\calA_j = \max \{ x_j^S ~|~ S \mbox{ active}\}$),
so that $w(X) = W\cdot X + E[c \cdot x^\calA]$.

We implement \ref{alg:submodular} for the \prob{Submodular-Cost Covering} instance $(w,\calC')$.
In an iteration of the algorithm for a constraint $S$ on $x^S$, the algorithm computes $\beta$ as follows.
Recall that the variables in $X$ being increased (to satisfy $x^S\in S$)
are $x^S_j$ for $j\in\vars(S)$.  
The derivative of $w(X)$ with respect to $x^S_j$ is
\begin{eqnarray*}
c'_j 
&=&
W^S_j + c_j \Pr[x^S_j \mbox{ determines } x^\calA_j]
\\&=&
W^S_j + c_j p_S\prod \big\{1 - p_{R} ~\big|~  x^{R}_j > x^S_j,~ j\in\vars(R)\big\}.
\end{eqnarray*}
The derivative will be constant
(that is, $w(X)$ will be linear in $x^S$)
until $x^S_j$ reaches its next breakpoint
$t_j = \min\{ x^R_j ~|~ x^{R}_j > x^S_j, j\in\vars(R)\}$.
Define $\beta_t = \min\{(t_j-x^S_j)c'_j ~|~ j\in\vars(S)\}$ to be
the minimum cost to bring any $x^S_j$ to its next breakpoint.

Let $w'$ be the vector defined above (the gradient of $w$ with respect to $x^S$).
Let $\beta'$ be the step size that the algorithm in Thm.~\ref{thm:CMIP} would compute
given the linear cost $w'$.
That is, that it would compute in an iteration for constraint $x^S\in S$
given the \prob{CMIP-UB} instance $(w',\{S\})$ and the current $x^S$.

The algorithm here computes $\beta_t$ and $\beta'$ as defined above,
then takes the step size $\beta$ to be $\beta= \min(\beta_t,\beta')$.
This $\beta$ is a valid lower bound on $\rescost X S$,
because $\beta_t$ is the minimum cost to bring any $x^S_j$ to its next breakpoint,
while $\beta'\le\rescp {x^S} (S)$ is a lower bound on the cost to satisfy $S$
without bringing any $x^S_j$ to a breakpoint.
Thus, by Thm.~\ref{thm:generic}, 
this algorithm computes a $\degree$-approximation.

The algorithm is as follows.
It considers the constraints in any order.
For each constraint $S$, it does iterations for that $S$,
with step size $\beta$ defined above, until $S$ is satisfied.

\begin{lemma}[\bf iterations]
For each $S\in\calC$,
the algorithm does at most $|\vars(S)|(\widehat\degree+2)$ iterations for $S$.
\end{lemma}
\begin{Proof}
An iteration may cause some $x^S_j$ to reach its next breakpoint $t_j$.
By inspection of the breakpoints $t_j$,
each $x^S_j$ can cross at most $\widehat\degree$ breakpoints
(one for each constraint $R$ on $x_j$ in the original instance).
Thus, there are at most $|\vars(S)|\widehat\degree$ such iterations.
In each remaining iteration the step size $\beta$ equals the step size $\beta'$
from the algorithm in Thm.~\ref{thm:CMIP}.
Following the proof of Lemma~\ref{lemma:hitting} in Thm.~\ref{thm:CMIP}, 
there are at most $2|\vars(S)|$ such iterations.
(In each such iteration,
either some variable $x_j^S$ reaches its upper bound $u_j$ for the first time,
or the constraint $x_j^S\in S^h$ is satisfied for the current relaxation $S^h$ of $S$.
By inspection, $S^h$ depends only on the current $x^S$ and the constraint $S$,
and not on the cost function $w'$.  Thus, as in the proof of Lemma~\ref{lemma:hitting},
after an iteration for $S$ where the current $S^h$ is satisfied, in the next iteration, $h$
will be at least one larger.   That can happen at most $|\vars(S)|$ times.)
\end{Proof}

To complete the proof of Thm.~\ref{thm:2stage}, we prove that
algorithm can be implemented to take time $O(N\widehat\degree\log\degree)$,
or, if $p=\mathbf 1$, time $O(N\log\degree)$.

As the algorithm does iterations for $S$, the algorithm maintains the data structures
described at the end of the proof of Thm.~\ref{thm:CMIP}, with the following adjustments.
When some $x^S_j$ reaches its next breakpoint and $w'_j$ increases,
the algorithm
\begin{itemize}
\item raises $x^b_j$ to maintain the invariant $x_j = x^b_j + \tau/w'_j$;
\item updates the derivative $d$ to account for the change in the term $a_j/c_j$ (if present in the derivative), and
\item updates the values for key $j$ in the three heaps (where present).
\end{itemize}
By inspection of the proof of Thm.~\ref{thm:CMIP}, these adjustments
are enough to maintain the data structures correctly throughout all iterations for $S$.
The updates take $O(\log \degree)$ time per breakpoint.
Thus, the total time for the adjustments is
$O(\sum_S |\vars(S)|\widehat\degree \log\degree)$,
which is $O(N\widehat\degree\log\degree)$.

To compute $\beta_t$ in each iteration, the algorithm does the following.
As it is doing iterations for a particular constraint $S$,
recall that $\tau$ is the sum of the $\beta$'s for $S$ so far
(from the proof of Thm.~\ref{thm:CMIP}).
The algorithm maintains a fourth heap
containing values $\{\tau + (t_j-x^S_j) w'_j ~|~ j\in \vars(S)\}$
(the values in the definition of $\beta_t$, plus $\tau$).
Then $\beta_t$ is the minimum value in this heap, minus $\tau$.

Then $x^S_j$ reaches a breakpoint (and $w'_j$ changes)
if and only if $\beta = \beta_t$ and key $j$ has minimum value in this heap.
When that happens, the algorithm finds the next breakpoint $t'_j$ for $j$
(as described in the next paragraph)
and updates $j$'s value in the fourth heap.
The total time spent maintaining the fourth heap is $O(\log\degree)$ per breakpoint,
$O(\sum_S\sum_{j\in\vars(S)} \widehat\degree \log\degree) = O(N\widehat\degree\log\degree)$.

The algorithm computes the breakpoints $t_j$ efficiently as follows.
Throughout the entire computation (not just the iterations for $S$),
the algorithm maintains, for each $j$, an array of $j$'s variables in $X$, 
that is,  $\{x_j^R ~|~ R\in\calC, j\in\vars(R)\}$,
sorted by the variables' current values (initially all 0).
Then $t_j$ is the value of the first $x_j^R$ after $x_j^S$ in $j$'s list.
When $x_j^S$ reaches its breakpoint $t_j$
(detected as described in the previous paragraph),
the algorithm updates the list order by swapping $x_j^S$
with the $x_j^R$ following it in the list (the one with value $t_j$).
The next breakpoint is then the value of the variable $x_j^{R'}$
that was after $x_j^R$ and is now after $x_j^S$.
The time spent computing breakpoints in this way is proportional
to the total number of swaps, which is proportional to the
total number of breakpoints, which is at most
$\sum_{S}\sum_{j\in\vars(S)} \widehat\degree = N\widehat\degree$.

This concludes the proof for the general case.
\smallskip

When $p=\mathbf 1$, note that in this case the product in the equation for $c_j'$
is 1 if  $x^S_j = \max_R x^R_j$ and 0 otherwise.
So each constraint $S$ has at most one breakpoint per variable,
and the total time for the adjustments above
reduces to $O(\sum_S |\vars(S)|\log\degree) = O(N \log \degree)$.
As in the proof of Thm.~\ref{thm:CMIP}, 
the remaining operations also take $O(N \log \degree)$ time.

This concludes the proof of the theorem.
\end{Proof}

\section*{Acknowledgements}
The authors gratefully acknowledge Marek Chrobak for useful discussions,
and two anonymous reviewers for careful and constructive reviews
that helped improve the presentation.

This work was partially supported by National Science Foundation (NSF) grants CNS-0626912 and CCF-0729071.

{
\bibliographystyle{plain}\small
\bibliography{bib}

\begin{thebibliography}{10}

\bibitem{Albers02On-Generalized}
S.~Albers.
\newblock On generalized connection caching.
\newblock {\em Theory of Computing Systems}, 35(3):251--267, 2002.

\bibitem{Bansal07A-Primal-Dual}
N.~Bansal, N.~Buchbinder, and J.~S. Naor.
\newblock A primal-dual randomized algorithm for weighted paging.
\newblock In {\em the forty-third IEEE symposium on Foundations Of Computer
  Science}, pages 507--517, 2007.

\bibitem{Bansal08RandomizedCompetitive}
N.~Bansal, N.~Buchbinder, and J.~S. Naor.
\newblock Randomized competitive algorithms for generalized caching.
\newblock In {\em the fourtieth ACM Symposium on Theory Of Computing}, pages
  235--244, 2008.

\bibitem{Bar-Yehuda00One-for-the-price}
R.~Bar-Yehuda.
\newblock One for the price of two: A unified approach for approximating
  covering problems.
\newblock {\em Algorithmica}, 27(2):131--144, 2000.

\bibitem{Bar-Yehuda04Local}
R.~Bar-Yehuda, K.~Bendel, A.~Freund, and D.~Rawitz.
\newblock {Local ratio: a unified framework for approximation algorithms}.
\newblock {\em ACM Computing Surveys}, 36(4):422--463, 2004.

\bibitem{Bar-Yehuda81A-Linear-Time}
R.~Bar-Yehuda and S.~Even.
\newblock {A linear-time approximation algorithm for the Weighted Vertex Cover
  problem}.
\newblock {\em Journal of Algorithms}, 2(2):198--203, 1981.

\bibitem{Bar-Yehuda85A-local-ratio}
R.~Bar-Yehuda and S.~Even.
\newblock {A local-ratio theorem for approximating the Weighted Vertex Cover
  problem}.
\newblock {\em Annals of Discrete Mathematics}, 25(27-46):50, 1985.

\bibitem{Bar-Yehuda01Efficient}
R.~Bar-Yehuda and D.~Rawitz.
\newblock Efficient algorithms for integer programs with two variables per
  constraint.
\newblock {\em Algorithmica}, 29(4):595--609, 2001.

\bibitem{Bar-Yehuda05On-the-equivalence}
R.~Bar-Yehuda and D.~Rawitz.
\newblock On the equivalence between the primal-dual schema and the local-ratio
  technique.
\newblock {\em SIAM Journal on Discrete Mathematics}, 19(3):762--797, 2005.

\bibitem{Bertsimas98Rounding}
D.~Bertsimas and R.~Vohra.
\newblock Rounding algorithms for covering problems.
\newblock {\em Mathematical Programming}, 80(1):63--89, 1998.

\bibitem{Borodin05How-well}
A.~Borodin, D.~Cashman, and A.~Magen.
\newblock How well can primal-dual and local-ratio algorithms perform?
\newblock In {\em the thirty-second International Colloquium on Automata,
  Languages and Programming}, 2005.

\bibitem{Borodin11How-well}
A.~Borodin, D.~Cashman, and A.~Magen.
\newblock How well can primal-dual and local-ratio algorithms perform?
\newblock {\em ACM Transactions on Algorithms}, 7(3):29, 2011.

\bibitem{Buchbinder05OnLine}
N.~Buchbinder and J.~S. Naor.
\newblock Online primal-dual algorithms for covering and packing problems.
\newblock In {\em the thirteenth European Symposium on Algorithms}, volume 3669
  of {\em Lecture Notes in Computer Science}, pages 689--701. 2005.

\bibitem{Buchbinder09Online}
N.~Buchbinder and J.~S. Naor.
\newblock Online primal-dual algorithms for covering and packing problems.
\newblock {\em Mathematics of Operations Research}, 34(2):270--286, 2009.

\bibitem{Cao97Cost-aware}
P.~Cao and S.~Irani.
\newblock Cost-aware www proxy caching algorithms.
\newblock In {\em the 1997 USENIX Symposium on Internet Technology and
  Systems}, pages 193--206, 1997.

\bibitem{Carr00Strengthening}
R.~D. Carr, L.~K. Fleischer, V.~J. Leung, and C.~A. Phillips.
\newblock Strengthening integrality gaps for capacitated network design and
  covering problems.
\newblock In {\em the eleventh ACM-SIAM Symposium On Discrete Algorithms},
  pages 106--115, 2000.

\bibitem{Chrobak91New-results}
M.~Chrobak, H.~Karloff, T.~Payne, and S.~Vishwanathan.
\newblock New results on server problems.
\newblock {\em SIAM Journal on Discrete Mathematics}, 4(2):172--181, 1991.

\bibitem{Chudak07Efficient}
F.~A. Chudak and K.~Nagano.
\newblock {Efficient solutions to relaxations of combinatorial problems with
  submodular penalties via the Lov\'{a}sz extension and non-smooth convex
  optimization}.
\newblock In {\em the eighteenth ACM-SIAM Symposium On Discrete Algorithms},
  pages 79--88, 2007.

\bibitem{Chvatal79GreedySetCover}
V.~Chv\'{a}tal.
\newblock A greedy heuristic for the set-covering problem.
\newblock {\em Mathematics of Operations Research}, 4:233--235, 1979.

\bibitem{Cohen1999ConnectionCaching}
E.~Cohen, H.~Kaplan, and U.~Zwick.
\newblock Connection caching.
\newblock {\em In the thirty-first ACM Symposium on Theory Of Computing}, pages
  612 -- 621, 1999.

\bibitem{Cohen2000ConnectionCaching}
E.~Cohen, H.~Kaplan, and U.~Zwick.
\newblock Connection caching under various models of communication.
\newblock In {\em the twelfth ACM Symposium on Parallel Algorithms and
  Architectures}, pages 54 -- 63, 2000.

\bibitem{Cohen2003ConnectionCaching}
E.~Cohen, H.~Kaplan, and U.~Zwick.
\newblock Connection caching: Model and algorithms.
\newblock {\em Journal of Computer and System Sciences}, 67(1):92--126, 2003.

\bibitem{dilley99enhancement}
J.~Dilley, M.~Arlitt, and S.~Perret.
\newblock Enhancement and validation of {S}quid's cache replacement policy.
\newblock In {\em fourth International Web Caching Workshop}, 1999.

\bibitem{Dinur2005OnTheHardness}
I.~Dinur and S.~Safra.
\newblock On the hardness of approximating minimum vertex cover.
\newblock {\em Annals of Mathematics}, 162:439--485, 2005.

\bibitem{Fiat91Competitive}
A.~Fiat, R.~M. Karp, M.~Luby, L.~A. McGeoch, D.~D. Sleator, and N.~E. Young.
\newblock Competitive paging algorithms.
\newblock {\em Journal of Algorithms}, 12:685--699, 1991.

\bibitem{Khuller07Greedy}
T.~Gonzales, editor.
\newblock {\em Approximation Algorithms and Metaheuristics}, chapter 4 (Greedy
  Methods).
\newblock Taylor and Francis Books (CRC Press), 2007.

\bibitem{Hall86A-fast}
N.~G. Hall and D.~S. Hochbaum.
\newblock A fast approximation algorithm for the multicovering problem.
\newblock {\em Discrete Applied Mathematics}, 15(1):35--40, 1986.

\bibitem{Halldorsson1994GreedIsGood}
M.~M. Halld\'{o}rsson and J.~Radhakrishnan.
\newblock Greed is good: Approximating independent sets in sparse and
  bounded-degree graphs.
\newblock In {\em the twenty-sixth ACM Symposium on Theory Of Computing}, pages
  439--448, 1994.

\bibitem{Halldorsson1997GreedIsGood}
M.~M. Halld\'{o}rsson and J.~Radhakrishnan.
\newblock Greed is good: Approximating independent sets in sparse and
  bounded-degree graphs.
\newblock {\em Algorithmica}, 18(1):145--163, 1997.

\bibitem{Halperin2002Improved}
E.~Halperin.
\newblock {Improved approximation algorithm for the Vertex Cover problem in
  graphs and hypergraphs}.
\newblock {\em SIAM Journal on Computing}, 31(5):1608--1623, 2002.

\bibitem{Hastad2001SomeOptional}
J.~H\.{a}stad.
\newblock Some optimal inapproximability results.
\newblock {\em Journal of the ACM}, 48(4):798--859, 2001.

\bibitem{Hayrapetyan2005NetworkDesign}
A.~Hayrapetyan, C.~Swamy, and \'{E}. Tardos.
\newblock Network design for information networks.
\newblock In {\em the sixteenth ACM-SIAM Symposium On Discrete Algorithms},
  pages 933 -- 942, 2005.

\bibitem{Hochbaum82Approximation}
D.~S. Hochbaum.
\newblock {Approximation algorithms for the Set Covering and Vertex Cover
  problems}.
\newblock {\em SIAM Journal on Computing}, 11:555--556, 1982.

\bibitem{Hochbaum83BoundsSetCover}
D.~S. Hochbaum.
\newblock {Efficient bounds for the Stable Set, Vertex Cover, and Set Packing
  problems}.
\newblock {\em Discrete Applied Mathematics}, 6:243--254, 1983.

\bibitem{hochbaum1996aan}
D.~S. Hochbaum.
\newblock {\em Approximation algorithms for NP-hard problems}.
\newblock PWS Publishing, 1996.

\bibitem{iwata2009submodular}
S.~Iwata and K.~Nagano.
\newblock Submodular function minimization under covering constraints.
\newblock In {\em the fiftieth IEEE Symposium on Foundations of Computer
  Science}, pages 671--680, 2009.

\bibitem{Johnson73SetCover}
D.~S. Johnson.
\newblock Approximation algorithms for combinatorial problems.
\newblock In {\em the fifth ACM Symposium on Theory Of Computing}, pages 38 --
  49, 1973.

\bibitem{Johnson74SetCover}
D.~S. Johnson.
\newblock Approximation algorithms for combinatorial problems.
\newblock {\em Journal of Computer and System Sciences}, 9(3):256--278, 1974.

\bibitem{Karlin1988CompetitiveSnoopy}
A.~R. Karlin, M.~S. Manasse, L.~Rudolph, and D.~D. Sleator.
\newblock Competitive snoopy caching.
\newblock {\em Algorithmica}, 3:77 -- 119, 1988.

\bibitem{Khot2008VertexCover}
S.~Khot and O.~Regev.
\newblock {Vertex Cover might be hard to approximate to within 2-$\eps$}.
\newblock {\em Journal of Computer and System Sciences}, 74:335--349, 2008.

\bibitem{Kolliopoulos05Approximation}
S.~G. Kolliopoulos and N.~E. Young.
\newblock Approximation algorithms for covering/packing integer programs.
\newblock {\em Journal of Computer and System Sciences}, 71(4):495--505, 2005.

\bibitem{Koufogiannakis09Distributed}
C.~Koufogiannakis and N.~E. Young.
\newblock Distributed and parallel algorithms for weighted vertex cover and
  other covering problems.
\newblock In {\em the twenty-eighth ACM symposium on Principles of Distributed
  Computing}, pages 171--179, 2009.

\bibitem{Koufogiannakis09DistributedPacking}
C.~Koufogiannakis and N.~E. Young.
\newblock Distributed fractional packing and maximum weighted b-matching via
  tail-recursive duality.
\newblock In {\em the twenty-third International Symposium on Distributed
  Computing}, pages 221--238, 2009.

\bibitem{koufogiannakis2009greedy}
C.~Koufogiannakis and N.~E. Young.
\newblock Greedy $\delta$-approximation algorithm for covering with arbitrary
  constraints and submodular cost.
\newblock In {\em the thirty-sixth International Colloquium on Automata,
  Languages, and Programming}, volume 5555 of {\em Lecture Notes in Computer
  Science}, pages 634--652. 2009.

\bibitem{Koufogiannakis11Distributed}
C.~Koufogiannakis and N.~E. Young.
\newblock Distributed algorithms for covering, packing and maximum weighted
  matching.
\newblock {\em Distributed Computing}, 24:45--63, 2011.

\bibitem{Kuhn06The-price}
F.~Kuhn, T.~Moscibroda, and R.~Wattenhofer.
\newblock The price of being near-sighted.
\newblock {\em In the seventeenth ACM-SIAM Symposium On Discrete Algorithm},
  pages 980--989, 2006.

\bibitem{Lotker08RentLeaseBuy}
Z.~Lotker, B.~Patt-Shamir, and D.~Rawitz.
\newblock Rent, lease or buy: Randomized algorithms for multislope ski rental.
\newblock In {\em the twenty-fifth Symposium on Theoretical Aspects of Computer
  Science}, pages 503 -- 514, 2008.

\bibitem{Lovasz75SetCover}
L.~Lov\'{a}sz.
\newblock On the ratio of optimal integral and fractional covers.
\newblock {\em Discrete Math}, 13(4):383--390, 1975.

\bibitem{Mcgeoch91AStrongly}
L.~A. McGeoch and D.~D. Sleator.
\newblock A strongly competitive randomized paging algorithm.
\newblock {\em Algorithmica}, 6(1):816--825, 1991.

\bibitem{Monien1985RamseyNumbers}
B.~Monien and E.~Speckenmeyer.
\newblock {Ramsey numbers and an approximation algorithm for the Vertex Cover
  problem}.
\newblock {\em Acta Informatica}, 22:115--123, 1985.

\bibitem{Orlin2007AFasterStrongly}
J.~B. Orlin.
\newblock A faster strongly polynomial time algorithm for submodular function
  minimization.
\newblock {\em In the twelfth conference on Integer Programming and
  Combinatorial Optimization}, pages 240--251, 2007.

\bibitem{Orlin2009AFasterStrongly}
J.~B. Orlin.
\newblock A faster strongly polynomial time algorithm for submodular function
  minimization.
\newblock {\em Mathematical Programming}, 118(2):237--251, 2009.

\bibitem{papadimitriou1993linear}
C.~H. Papadimitriou and M.~Yannakakis.
\newblock Linear programming without the matrix.
\newblock In {\em the twenty-fifth ACM Symposium on Theory of Computing}, pages
  121--129, 1993.

\bibitem{Pritchard09Approximability}
D.~Pritchard.
\newblock Approximability of sparse integer programs.
\newblock In {\em the seventeenth European Symposium on Algorithms}, volume
  5757 of {\em Lecture Notes in Computer Science}, pages 83--94. 2009.

\bibitem{Pritchard11Approximability}
D.~Pritchard and D.~Chakrabarty.
\newblock Approximability of sparse integer programs.
\newblock {\em Algorithmica}, 61(1):75--93, 2011.

\bibitem{Raghavan94Memory}
P.~Raghavan and M.~Snir.
\newblock Memory versus randomization in on-line algorithms.
\newblock {\em IBM Journal of Research and Development}, 38(6):683--707, 1994.

\bibitem{ravi2006hua}
R.~Ravi and A.~Sinha.
\newblock Hedging uncertainty: Approximation algorithms for stochastic
  optimization problems.
\newblock {\em Mathematical Programming}, 108(1):97--114, 2006.

\bibitem{shmoys2004soa}
D.~Shmoys and C.~Swamy.
\newblock Stochastic optimization is (almost) as easy as deterministic
  optimization.
\newblock In {\em the forty-fifth IEEE symposium on Foundations Of Computer
  Science}, pages 228--237, 2004.

\bibitem{Sleator85Amortized}
D.~D. Sleator and R.~E. Tarjan.
\newblock Amortized efficiency of list update and paging rules.
\newblock {\em Communications of the ACM}, 28(2):202--208, 1985.

\bibitem{Srinivasan99Improved}
A.~Srinivasan.
\newblock Improved approximation guarantees for packing and covering integer
  programs.
\newblock {\em SIAM Journal on Computing}, 29:648--670, 1999.

\bibitem{Srinivasan01NewApproaches}
A.~Srinivasan.
\newblock New approaches to covering and packing problems.
\newblock In {\em the twelveth ACM-SIAM Symposium On Discrete Algorithms},
  pages 567 -- 576, 2001.

\bibitem{vazirani2001aa}
V.~V. Vazirani.
\newblock {\em Approximation algorithms}.
\newblock Springer, 2001.

\bibitem{young1991online}
N.~E. Young.
\newblock On-line caching as cache size varies.
\newblock In {\em the second ACM-SIAM Symposium On Discrete Algorithms}, pages
  241--250, 1991.

\bibitem{Young94The-k-Server}
N.~E. Young.
\newblock The k-server dual and loose competitiveness for paging.
\newblock {\em Algorithmica}, 11:525--541, 1994.

\bibitem{Young98Online}
N.~E. Young.
\newblock On-line file caching.
\newblock In {\em the twelveth ACM-SIAM Symposium On Discrete Algorithms},
  pages 82--86, 1998.

\bibitem{Young02On-Line}
N.~E. Young.
\newblock On-line file caching.
\newblock {\em Algorithmica}, 33(3):371--383, 2002.

\end{thebibliography}
}

\section*{Appendix}

\paragraph{Proof of Observation~\ref{obs:canonical} (reduction to canonical form).}
Here is the reduction:
Let $(c,U,\calC)$ be any instance of \prob{Submodular-Cost Covering}.
Construct its canonical form $(c,\calC')$ as follows.
First, assume without loss of generality that $\min U_j = 0$ for each $j$.
(If not, let $\ell_j = \min U_j$,
then apply the translation $x \leftrightarrow x' + \ell$ to the cost and feasible region:
rewrite the cost $c(x)$ as $c'(x') = c(x'+\ell)$;
rewrite each constraint ``$x\in S$'' as ``$x' \in S-\ell$'';
replace each domain $U_j$ by $U'_j = U_j - \ell_j$.)

Next, define $\mu_j(x) = \max\{\alpha \in U_j ~|~ \alpha \le x_j\}$
(that is, $\mu(x)$ is $x$ with each coordinate lowered into $U_j$).
For each constraint $S$  in $\calC$,
put a corresponding constraint ``$\mu(x) \in S$'' in $\calC'$.
The new constraint is closed upwards and closed under limit because $S$ is
and $\mu$ is non-decreasing.
It is not hard to verify that any solution $x$ to the canonical instance $(c,\calC')$
gives a corresponding solution $\mu(x)$ to the original instance $(c,U,\calC)$,
and that this reduction preserves $\degree$-approximation.
\hfill\qed



\end{document}